\documentclass[11pt,a4paper]{article}
\usepackage{jheppub_kim}
\usepackage{longtable}

\usepackage{epsfig}
\usepackage{amsmath}
\usepackage{graphicx}
\usepackage{hyperref}
\usepackage{epsfig,multicol,bbm}
\usepackage{lmodern}
\usepackage[T1]{fontenc}
 \usepackage{graphics}
 \usepackage[scriptsize,nooneline,hang]{caption}
\usepackage[hang,nooneline,scriptsize]{subfigure}
\usepackage{array}
\usepackage{makecell}
\usepackage{verbatim}
\begin{document}

\title{Astrophysical and electromagnetic emissivity properties of black holes surrounded by a quintessence type exotic fluid in the Scalar-Vector-Tensor Modified Gravity}

\author[a,b,c1]{Haidar Sheikhahmadi,}
\affiliation[a]{School of Astronomy, Institute for Research in Fundamental Sciences (IPM),\\
  P. O. Box 19395-5531, Tehran, Iran,}
\affiliation[b]{Center for Space Research, North-West University, Potchefstroom,  South Africa,}
\affiliation[c]{Canadian Quantum Research Center 204-3002 32 Avenue Vernon, British Columbia V1T 2L7, Canada}

\author[d]{Saheb Soroushfar,}
\affiliation[d]{Faculty of Technology and Mining, Yasouj University, Choram 75761-59836, Iran}

\author[e]{S. N. Sajadi,}
\affiliation[e]{School of Physics, Institute for Research in Fundamental Sciences (IPM), P. O. Box 19395-5531, Tehran, Iran}

\author[f,g,h]{Tiberiu Harko}
\affiliation[f]{Faculty of Physics, Babes-Bolyai University, Kogalniceanu Street,
400084 Cluj-Napoca, Romania,}

\affiliation[g]{Department of Theoretical Physics, National Institute of Physics
and Nuclear Engineering (IFIN-HH), Bucharest, 077125 Romania,}

\affiliation[h] {Astronomical Observatory, 19 Ciresilor Street, 400487 Cluj-Napoca, Romania,}


\emailAdd{h.sh.ahmadi@gmail.com;h.sheikhahmadi@ipm.ir}
\emailAdd{saheb.soroushfar@gmail.com}
\emailAdd{naseh.sajadi@gmail.com}
\emailAdd{tiberiu.harko@aira.astro.ro }

\begin{abstract}
{ The astrophysical consequences of the presence of a quintessence scalar field on the evolution of the horizon and on the accretion disk surrounding a static black hole, in the Scalar-Vector-Tensor version of Modified Gravity (MOG), are investigated.  The positions of the stable circular orbits of the massive test particles, moving around the central object, are obtained from the extremum of the effective potential.  Detailed calculations are also presented to investigate the light deflection, shadow and Shapiro effect for such a black hole. The electromagnetic properties of the accretion disks that form around such black holes are considered in detail. The energy flux and efficiency parameter are estimated analytically and numerically. A comparison with the disk properties in Schwarzschild geometry is also performed. The quantum properties of the black hole are also considered, and the Hawking temperature and the mass loss rate due to the Hawking radiation are considered. The obtained results may lead to the possibility of direct astrophysical tests of black hole type objects in modified gravity theories.}
\end{abstract}

\maketitle

\section{Introduction}
\label{Intro}

According to the $\Lambda $ cold dark matter, $\Lambda $CDM, model, the material content of the Universe
mainly consists of two basic components, dark energy, and dark matter, respectively \cite{Peebles:2002gy,Padmanabhan:2002ji}. To explain the observed late-time cosmic acceleration of the Universe \cite{Riess:1998cb,Perlmutter:1998np,Hanany:2000qf}, one can consider either a fundamental
cosmological constant $\Lambda $, which, for example, could be interpreted as an intrinsic
geometric property of the space-time, or a dark energy, an ambiguous  fluid component, of yet unknown physical origin, which spreads throughout the whole Universe, and which
would mimic a cosmological constant. Currently, one of the main dark energy scenarios is
based on the so-called \textit{quintessence} fields \cite{Wetterich:1987fm,Peebles:1987ek,Ratra:1987rm,Caldwell:1997ii}. In this model, the cosmological constant, $\Lambda,$  is considered as a variable term, and it's evolution and properties can be explained by considering a scalar field with negative pressure, representing the dark energy \cite{Amendola:1999er}.

In cosmological studies the quintessence model faces some drawbacks for providing a good concordance between theory and observations, especially in local scales. These theoretical problems led to the introduction of some extended scalar-tensor models, as chameleon, and Brans-Dicke chameleon models \cite{Brans:1961sx,Khoury:2003rn,Khoury:2003aq,Saaidi:2011zza,Sheikhahmadi:2018aux,Rabiei:2015pha,Aghamohammadi:2013eja}.
 Additionally, the explanation of the accelerated expansion phase itself in a consistent quantum model of gravity is still a challenge \cite{Kiselev:2002dx}. The origin of this drawback goes back to the so-called outer horizon problem. In other word, this problem is related to the impossibility of introducing an observable and physical S-matrix for describing the asymptotic in and out states.

In the following we will not investigate the cosmological properties, with their advantages and disadvantages, of the quintessence models.
Instead of the large scale cosmological studies we will consider the effects of the quintessence fields on the behaviour of the compact objects, as, for instance, the black holes.

Interestingly enough, in \cite{Kiselev:2002dx} it was shown that in the presence of the scalar field and  under some simple assumptions for the energy-momentum tensor, an exact black hole solution for the Einstein equation can be obtained. One interesting property of this solution is that when one considers the interval $-1<\omega<-1/3$ for the equation of state parameter $\omega=p/\rho,$, then an outer horizon is formed.

Before these investigations, in \cite{GonzalezDiaz:2001ce,GonzalezDiaz:2002nt}  a solution based on a free quintessence model was found, which leads to a bare singularity, with no hair and no horizon, with the cosmic censorship conjecture not satisfied. Therefore this solution does not represents a black hole solution. Recently, many different classes of black holes models in the presence of quintessence fields have been investigated. For the rotating charged black holes one can see\cite{Larranaga:2017ttl}. For the black hole solutions originating from string theory we refer the readers to \cite{Toledo:2018hav}. The thermodynamics properties of the charged black holes in the presence of quintessence  have been investigated in \cite{Wei:2011za}. In this work, different conditions for the heat conductivity of the black holes' three horizons have been discussed in detail. Entropy calculations of the black holes in the presence of quintessence can be found in \cite{Mureika:2015sda}. For more details on the black holes in the presence of quintessence fields  see \cite{deOliveira:2018weu,Toledo:2019mlz,Ghosh:2017cuq}, and references therein. It deserves to also note that there are some models of black hole solutions that are able to remove the central singularity, namely, the regular black holes, and  the non-commutative black holes \cite{Sajadi:2021ilt,Hendi:2020knv,Sajadi:2017glu}, respectively.

Let us now briefly review the astrophysical and physical properties of the accretion disks. Usually, in astrophysics the term accretion is associated to the process of the attraction of matter, especially gaseous, by a central massive object. When this ingoing process happens, gravitational energy, and even momentum, are extracted from the central object. Obviously, the amount of such energy should be proportional to the ratio $M/R$, where $M$ denotes the mass, and $R$ denotes the radius of the central object, respectively. Therefore it can be concluded that the more massive and compact the central mass is, the larger amount of energy will flow in. Hence, from a phenomenological point of view, the accretion  into a black hole is one of the important and interesting phenomena in the astrophysical context. Its importance goes back to the fact that it indicates the transfer of gravitational energy into radiation, and leads to an explanation of the  properties of the most luminous objects in the Universe,  which recently have been observed directly by the Event Horizon telescope, EHT \cite{Goddi:2019ams}.   Obviously, the spherical symmetric accretion on a cental compact object is the simplest case one can visualize. For the first time such a system was studied in \cite{1952MNRAS.112..195B}, and it is well-known as the Bondi accretion. Bondi did show that this solution is unique, and consequently the flow becomes transonic crossing the sonic radius. Observations have shown that accretion processes are present in many astrophysical objects.

An important consequence of the accretion processes is that many compact objects are surrounded by accretion disk. Astrophysical phenomena related to accretion disks have been   studied extensively. Here we give some simple example of accretion disks.  Usually, the protostellar clouds formed of a molecular gas can form a protoplanetary disk around the host newly formed star. As the second example, we can refer to the binary systems, in which mass can be accreted to form an accretion disk due to the presence of strong stellar winds.

The third possibility for the formation of an accretion disk is perhaps the more interesting one, and recently it has been confirmed by direct observations  \cite{Goddi:2019ams}. It is related to the super massive black holes in active galactic nuclei, which can acquire mass by accreting matter from their vicinity. It is important to emphasize that in almost all high luminosity astrophysical objects there is a central mass surrounded by  an accretion disk. For a valuable discussion of the accretion disk properties of compact objects we refer the reader to \cite{PadmanabhanT}. The study of the properties and of the physics of the Newtonian accretion disks was initiated in the important and seminal paper \cite{1973A&A....24..337S}.

There are many  different proposals advanced to explain the rotation curves of galaxies, the dynamics of galaxy clusters, and  other astrophysical and cosmological phenomena. In one hand, as we have already mentioned above, the dark matter model, for the first time hypothesized by Zwicky \cite{Zwicky:1929zz}, has become an essential part of the standard $\Lambda$CDM cosmological paradigm. On the other hand, to explain the dark matter phenomenology, models based on modified gravity have also been proposed. Fore example, some models, based on imposing a modification in Newtonian version of gravity, the so called Modified Newtonian dynamics, MOND \cite{Milgrom:1983ca} have been considered as alternatives to the standard dark matter model.

Another proposal for a modified gravity theory is the scalar-vector-tensor (S-V-T) modified gravity theory, or MOG \cite{Moffat:2005si}. This theoretical model adequately explains Solar System observations \cite{Moffat:2014asa}, the rotation curves of galaxies \cite{Moffat:2013sja}, the dynamics of galaxy clusters \cite{Brownstein:2007sr}, and the Cosmic Microwave Background predictions \cite{Moffat:2014bfa}. Recently, the physics of some black hole solutions in the MOG version of gravity has been investigated, and it has been shown that the well known exact solutions of general relativity can be reproduced properly \cite{Moffat:2015kva,Moffat:2014aja}. For instance, although the Reissner-Nordtr\"{o}m solution in the standard version of gravity theory (general relativity) corresponds to an electrically charged object, in MOG it can be considered as describing electrically neutral objects. In the MOG theory, a gravitational repulsive force, with an effective charge $Q=\sqrt{\alpha G_N}M$, can be introduced. The parameter $\alpha=-1+G/G_N$ is known as the MOG parameter, and it has to be determined from the observational constraints. The parameter $G_N$ is Newton's constant, and $G$ is the enhanced gravitational constant. Moreover, by $M$ we have denoted the total mass of the central object, a black hole, for example.

This physical solution for a MOG black hole motivated us to extend our investigation to study the properties of the black holes in the presence of a scalar field, that is, a  quintessence field, and to consider the properties of the compact objects in such an extension of the MOG theory. In the following we will cal, this solution of the MOG theory as a QMOG black hole. More exactly, we consider first the positions of the stable circular orbits of the massive test particles, moving around the central MOG type black hole, which are obtained from the extremum of the effective potential. Several important astrophysical effects are investigated in detail, including  the light deflection, the shadow and the Shapiro effect, respectively. The electromagnetic properties of the accretion disks are also analyzed in detail, and  the energy flux, the luminosity, and the efficiency parameter are obtained by using both analytical and numerical methods. For each quantities we perform a comparison with the disk properties in Schwarzschild geometry. The important quantum properties of the MOG black holes are analyzed, and the Hawking temperature, and the mass loss rate due to the Hawking radiation is obtained. The results presented in this study may lead to the possibility of the astrophysical tests of the black hole type objects in modified gravity theories, and of the modified gravity theories themselves.

This work is organised as follows. In Sec.~\ref{SVT_RN}, we write down the Reissner-Nordstr\"{o}m type black hole solution in the presence of a quintessence field in the S-V-T theory. Moreover, the dynamical properties of the motion of the massive particles in this geometry (effective potential, positions of the event horizon, and the radii of the marginally stable orbits) are analyzed in detail  The astrophysical properties of the QMOG black holes, like the deflection of light, the shadow of the black holes, and the Shapiro effects are studied in Section~\ref{sectastro}. Furthermore, in Sec.\eqref{Fluxetc...}, the flux, temperature and the efficiency of the accretion disks that form around QMOG black hole are obtained both analytically, and numerically. Finally, in Sec.~\ref{section6} we discuss and conclude our results.

\section{Kiselev type S-V-T black hole in the presence of an exotic fluid}\label{SVT_RN}

In the following Section we will first write down the metric of a static black hole in the S-V-T (MOG) version of gravity in the presence of a quintessence scalar field, aiming at investigating the implications of this modified gravity theory on the behaviour and properties of the massive astrophysical objects. Moreover, the horizon properties and the equations of motion of the massive test particles are analyzed in detail. In the next step, the dynamical properties of this compact, black hole type object, will be analyzed as well.

\subsection{General properties of the metric, and the structure of the horizons}

In this Section, we consider a static and spherically symmetric black hole, surrounded by a quintessence scalar field, in the MOG theory. Generally, the geometry of the static, spherically symmetric black hole is given by \cite{Kiselev:2002dx,Moffat:2015kva,Moffat:2014aja,Nandan:2016ksb,Yang:2016sjy,Abbas:2018ygc}
\begin{equation}\label{metric}
ds^{2}=-f(r)dt^{2}+f(r)^{-1}dr^{2}+r^{2}(d\theta^{2}+\sin^{2}\theta d\varphi^{2}).
\end{equation}

In the following we consider a Kiselev type black hole in MOG, with the metric function given by
\begin{equation}\label{metricfunction}
f(r)= 1-\frac{2G M}{r}+\frac{GQ^{2}}{r^{2}}-\frac{ \gamma G}{r^{3\epsilon+1}},
\end{equation}
where we have denoted $G=G_N(1+\alpha)$,  and $ M$  is the black hole mass, $Q=\sqrt{\alpha G_N}M$ is {\it the effective (non-electric) charge}, while $\alpha$  is the MOG parameter, $ \gamma$ is the quintessence parameter, and  $\epsilon$  is the equation of state parameter of the quintessence field. The allowed values for $\epsilon$ are in the range  $-1 <  \epsilon  < -1/3$. In the following we will call the black hole described by the metric (\ref{metricfunction}) as the QMOG black hole.

\begin{figure}[h]
	\centering
	\subfigure{
		\includegraphics[width=0.5\textwidth]{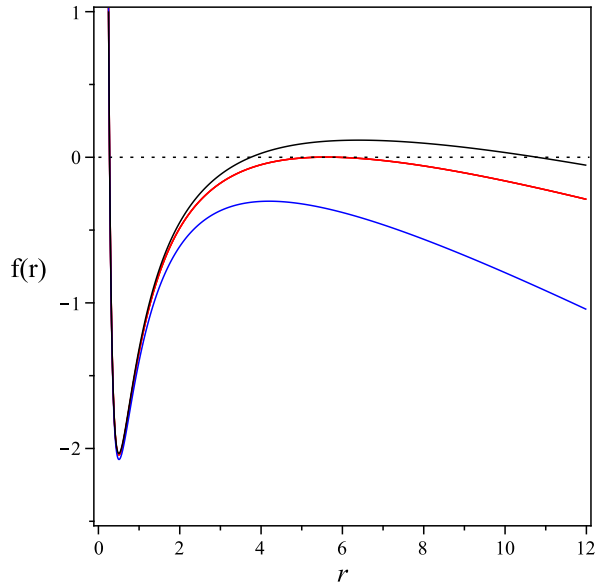}
	}
	\caption{The plots of $f(r)$  for $M=1,\epsilon=-2/3, G_{N}=1, \alpha=0.5$ and $\gamma=\textcolor{blue}{0.1},\textcolor{red}{0.045},0.058$.}
	\label{F(r)}
\end{figure}

In the case of $\epsilon=-1$, the metric (\ref{metricfunction}) reduces to the general relativistic black hole space-times, in the presence of a cosmological constant, i.e., to the  anti de-Sitter Schwarzschild geometry (for more details see \cite{Moffat:2015kva}). Moreover, in the limit of $\gamma= 0$, the metric reduces to an {\it effective Reissner–Nordström type black hole}, with a non-electric charge. For $\gamma = 0$, and  $Q=0,$, it reduces to the well-known Schwarzschild black hole. For the case  $ \epsilon = -2/3$, Eq.~\eqref{metric} describes an effective Reissner–Nordström-like black hole, surrounded by quintessence field in the S-V-T (MOG) version of gravity theories. In Fig.~\ref{F(r)} the behaviour of the metric coefficient  is illustrated, for various numerical values of the metric parameters.

It deserves here to note that Moffat \cite{Moffat:2015kva} obtained his MOG version of the black holes by neglecting  the energy momentum tensor of the matter sector, i.e., $T_{\mu\nu}^m=0$. Following \cite{Kiselev:2002dx}, and by introducing the matter sector, as a quintessence scalar field, the metric \eqref{metric} obtained accordingly. To obtain the metric one must decompose the energy momentum tensor, in the Einstein field equation $G_{\mu\nu}=-8\pi T_{\mu\nu}$, in two distinct components, i.e.,
\begin{equation}\label{Tmunu}
T_{\mu\nu}=T_{\mu\nu}^{q}+T_{\mu\nu}^{V}\,,
\end{equation}
where $T_{\mu\nu}^{q}$ stands for the quintessence part \cite{Kiselev:2002dx}, and $T_{\mu\nu}^{V}$ describes the vector field \cite{Moffat:2015kva}.
For the derivation of the metric one can find more details in \cite{Karakasis:2021rpn,Sotiriou:2015pka,Carneiro:2018url} as well.

It is important here to note that the Kiselev black holes can be interpreted as a solution of the Einstein gravitational field equations with the matter energy-momentum
tensor given by \cite{Visser},
\begin{equation}
T_t^t=T_r^r=\rho (r),
\end{equation}
and
\begin{equation}
T_\theta^\theta=T_\phi^\phi=-\frac{1}{2}\left(1+3\varepsilon\right)\rho,
\end{equation}
where $\varepsilon $ is the parameter of the equation of state. By taking the isotropic average over the angles of the components of the energy-momentum tensor we obtain the barotropic equation of state $p=\varepsilon\rho$. Hence, the interpretation of the Kiselev black hole as describing a black hole solution in the presence of a quintessence field is problematic, and this solution may be better interpreted as describing a black hole in the presence of an exotic matter source.

\subsection{Horizons of the QMOG black hole}

The horizon of the spacetime given by the metric (\ref{metric}) can be determined by setting  the condition $ g_{00}(r)=0 $, i.e.,
\begin{equation}
1-\frac{2 G M}{r}+\frac{G Q^{2}}{r^{2}}-\frac{G\gamma}{r^{3\epsilon+1}}=0.
\end{equation}
Consequently, we obtain for the position of the horizon the algebraic equation
\begin{equation}\label{Horizon}
r^{2}-2 G M r+G Q^{2}-G \gamma  r^{1-3\epsilon}=0 .
\end{equation}

For our purposes, and for mathematical convenience, we introduce the following dimensionless  parameters, defined as,
\begin{eqnarray}\label{tilde}
r_\star=\frac{\tilde{r}}{(1+\alpha)}=\frac{r}{(1+\alpha)G_N M},\\ \nonumber
 Q_\star=  \frac{\tilde{Q}}{\sqrt{1+\alpha}}=\frac{Q}{\sqrt{G_N(1+\alpha)} M},\\ \nonumber
~ \gamma_\star=(1+\alpha)^{2}{\tilde{\gamma}}=(1+\alpha)^{2}G_N^2 \gamma M\,.
\end{eqnarray}

For $ \epsilon = -2/3$, and by rearranging the parameters, Eq.\eqref{Horizon}, interestingly, takes the following form,
\begin{equation}\label{Horizon2}
\gamma_\star {r}_\star^{3}-{r}_\star^{2}+ 2{r}_\star-{{Q}_\star^2}=0~.
\end{equation}
Another interesting case corresponds to $ \epsilon = -\frac{1}{2} $. We have to note that for this value of $\epsilon$,  the dimensionless parameter $\tilde{\gamma}$ takes the form $\tilde{\gamma}=G_N^3 \gamma^2 M$. Accordingly, Eq.\eqref{Horizon} can be rearranged as follows,
\begin{equation}\label{Horizon22}
\sqrt{\gamma_\star }{r}_\star^{\frac{5}{2}}-{r}_\star^{2}+ 2{r}_\star-{{Q}_\star^2}=0.
\end{equation}

In the following we would like to investigate the effects of the MOG parameter $\alpha$ on the properties of the black holes. To this end, by taking into account the definitions of the effective charge, and of the enhanced gravitational constant,  one can rewrite Eqs.(\ref{Horizon2}) and (\ref{Horizon22}), respectively, in the alternative forms, namely,
\begin{equation}\label{Horizon2MOG}
\tilde{\gamma}\tilde{r}^{3}-\frac{\tilde{r}^{2}}{(1+\alpha)}+ 2\tilde{r}-{\alpha}=0~\,,
\end{equation}
and
\begin{equation}\label{Horizon22MOG}
\sqrt{\tilde{\gamma}(1+\alpha)}\tilde{r}^{\frac{5}{2}}-{\tilde{r}^{2}}+2(1+\alpha){{\tilde{r}}}-{\alpha(1+\alpha)}=0~,
\end{equation}
respectively.

{It is interesting to notice that there is another possibility of extending the range of $\epsilon$ by also considering the case $\epsilon>0$. From the asymptotically flat behaviour of such a configuration it can be concluded that, obviously, it would not obey the properties required for a quintessence field anymore. To make these things clearer, as an example, let's consider the value $\epsilon=2/3$. Then, the dimensionless equation \eqref{Horizon} can be reformulated as,
\begin{equation}\label{HorizonPlus}
\gamma_\star {r}_\star^{-3}-{{Q}_\star^2}{r}_\star^{-2}+ 2{r}_\star^{-1}-1=0~,
\end{equation}
where, for the positive values of $\epsilon$, i.e., $\epsilon=2/3$, one has $\bar{\gamma}=G_N^2 \gamma M^3$. Now, from Eq.(\ref{tilde}), and from the definition of $Q$, one can rewrite Eq.(\ref{HorizonPlus}) as follows,
\begin{equation}\label{HorizonPlusMOG}
\bar{\gamma}(1+\alpha)^{3}\tilde{r}^{-3}-\frac{\alpha}{1+\alpha}{\tilde{r}^{-2}}+\frac{2}{1+\alpha}\tilde{r}^{-1}-\frac{1}{(1+\alpha)^{2}}=0~.
\end{equation}
}
We will return to this matter, and discuss it in detail,  in the next Sections.

\subsection{Equations of motion, and the effective potential}


In this Section, to investigate the physical properties of the MOG black hole in the presence of a quintessence field (or an exotic matter source), we are going to study first the equations of motion of a massive test particle, and to derive the effective potential as well. It is well known that for a point, massive test particle, the Lagrangian $ \mathcal{L} $ of the motion in  the spacetime given by Eq.(\ref{metric}), can be written as,
\begin{equation}
\mathcal{L}=\frac{1}{2}g_{\mu\nu}\frac{dx^\mu}{ds}\frac{dx^\nu}{ds}\,,
\end{equation}
Accordingly, in the equatorial plane, the conserved energy $E$ and the angular momentum $L$ can be obtained according to,
\begin{equation}
E=g_{tt}\frac{dt}{ds}=(1-\frac{2 G M}{r}+\frac{G Q^{2}}{r^{2}}-\frac{G\gamma}{r^{3\epsilon+1}})\frac{dt}{ds}\,,
\end{equation}
and
\begin{equation}
L=g_{\varphi\varphi}\frac{d\varphi}{ds}=r^{2}\frac{d\varphi}{ds}\,,
\end{equation}
respectively.  Plugging the above equations all together, the geodesic equations are expressed in the form,
\begin{equation}\label{dr/ds}
\left(\frac{dr}{ds}\right)^2=E^2-\Bigg(1-\frac{2G  M}{r}+\frac{G Q^{2}}{r^{2}}-\frac{G \gamma}{r^{3\epsilon+1}}\Bigg)\left(1+\frac{L^2}{r^2}\right)\,,
\end{equation}
\begin{eqnarray}\label{dr/dphi}
\left(\frac{dr}{d\varphi}\right)^2=\frac{r^4}{L^2}\Bigg[E^2-\left(1-\frac{2 G M}{r}+\frac{G Q^{2}}{r^{2}}-\frac{G \gamma}{r^{3\epsilon+1}}\right)\left(1+\frac{L^2}{r^2}\right)\Bigg]\,,
\end{eqnarray}
\begin{eqnarray}\label{dr/dt}
\left(\frac{dr}{dt}\right)^2=\frac{1}{E^2}\Bigg(1-\frac{2G  M}{r}+\frac{G Q^{2}}{r^{2}}-\frac{G \gamma}{r^{3\epsilon+1}}\Bigg)^2\times\\\nonumber
\Bigg[E^2-\left(1-\frac{2G M}{r}+\frac{G Q^{2}}{r^{2}}-\frac{G \gamma}{r^{3\epsilon+1}}\right)(1+\frac{L^2}{r^2})\Bigg]\,.
\end{eqnarray}
 See \cite{Soroushfar:2015wqa,Soroushfar:2015dfz,Soroushfar:2016esy,Hoseini:2016nzw,Soroushfar:2016yea,Hoseini:2016tvu} for more details on the derivation of these equations.

 Eqs.(\ref{dr/ds})-(\ref{dr/dt}) give a complete description of the
dynamics of the massive test particles moving around the MOG black hole. By considering Eq.(\ref{dr/ds}), we can define an effective
potential of the motion as,
\begin{equation}\label{V}
V_{eff}=\Big(1-\frac{2 GM}{r}+\frac{GQ^{2}}{r^{2}}-\frac{G \gamma}{r^{3\epsilon+1}}\Big)\left(1+\frac{L^2}{r^2}\right).
\end{equation}

These definition allows us to compare our results with the standard form of the equations of motion in other geometries, and black hole solutions.

Now we are prepared to discuss the properties of the accretion disks that form around the central mass. We will begin this discussion in the next Subsection by considering the properties of the event horizons, and of the stable circular orbits for the QMOG black holes.

\subsection{Event horizons, and stable circular orbits of the QMOG black hole}\label{ThinAcc disk}

 The accretion disks usually form through a simple astrophysical mechanism. In an accretion disk, hot gas particles, carrying an electric charge, are moving in stable circular orbits around the central compact object, which in the present case we consider to be a black hole.
For the case $ \epsilon = -2/3 $, the specific energy $ \tilde{E} $, the specific angular momentum $ \tilde{L} $, and the angular velocity $ \tilde{\Omega} $, of the particles that move in a circular orbit can be written as,
\begin{equation}\label{E}
\tilde{E}=\dfrac{\tilde{g}_{tt}}{\sqrt{\tilde{g}_{tt}-\tilde{g}_{\phi\phi}\tilde{\Omega}^{2}}}=\frac{\sqrt{2} \Bigg[\alpha  (\alpha +1)-(\alpha +1) \tilde{\gamma } \tilde{r}^3+\tilde{r}^2-2 (\alpha +1) \tilde{r}\Bigg]}{(\alpha +1) \tilde{r}^2 \sqrt{\frac{2}{\alpha +1}-\frac{6 \tilde{r}-4 \alpha }{\tilde{r}^2}-\tilde{\gamma} \tilde{ r}}},
\end{equation}
\begin{equation}\label{L}
\tilde{L}=\dfrac{\tilde{g}_{\phi\phi}{\tilde{\Omega}}}{\sqrt{\tilde{g}_{tt}-\tilde{g}_{\phi\phi}\tilde{\Omega}^{2}}}= \frac{ \sqrt{{-2 \alpha -\gamma  \tilde{r}^3+2 \tilde{r}}}}{\sqrt{\frac{2}{\alpha +1}-\frac{6 \tilde{r}-4 \alpha }{\tilde{r}^2}-\tilde{\gamma} \tilde{ r}}},
\end{equation}
and,
\begin{equation}\label{Om}
\tilde{\Omega}=\sqrt{\dfrac{\tilde{g}_{tt,\tilde{r}}}{\tilde{g}_{\phi\phi,\tilde{r}}}}= \frac{\sqrt{-2 \alpha -\tilde{\gamma}  \tilde{r}^3+2\tilde{ r}}}{\sqrt{2} \tilde{r}^2},
\end{equation}
respectively.

Using Eqs.(\ref{tilde}), (\ref{V}) and (\ref{L}), for the effective potential, and its derivatives, we obtain,
\begin{equation}\label{Vtilde}
\tilde{V}_{eff}=\dfrac{1}{\tilde{r}^2}\Bigg[-\tilde{\gamma}\tilde{r}^{3}+\frac{\tilde{r}^{2}}{(1+\alpha)}-2\tilde{r}+{\alpha}\Bigg]\left(1+\frac{\tilde{L}^2}{\tilde{r}^2}\right) ,
\end{equation}
\begin{equation}\label{Vtilde1}
\begin{array}{l}
\dfrac{d\tilde{V}_{eff}}{d\tilde{r}}= - \frac{1}{{(1 + \alpha ){r^5}}}\bigg[ - {\tilde L^2}\alpha \,\tilde \gamma \,{r^3} + \alpha \,\tilde \gamma {r^5} - {\tilde L^2}\tilde \gamma {r^3} + \tilde \gamma {r^5} + 4{\tilde L^2}{\alpha ^2}\\
  \quad\quad ~~~~~~ - 6{L^2}\alpha r + 2{\tilde L^2}{r^2} + 2{\alpha ^2}{r^2} - 2\alpha {r^3} + 4{\tilde L^2}\alpha  - 6{\tilde L^2}r + 2\alpha {r^2} - 2{r^3}\bigg]\,,
 \end{array}
 \end{equation}
 and
 \begin{eqnarray}\label{V2}
\begin{array}{l}
\dfrac{d^{2}\tilde{V}_{eff}}{d\tilde{r}^{2}} =\frac{2}{{\big(1 + \alpha \big){{\tilde {r}}^6}}} \times
\Bigg[ - {{\tilde L}^2}\tilde \gamma {\tilde{r}^3}(\alpha  + 1) + 10{{\tilde L}^2}{\alpha ^2} - 12{{\tilde L}^2}\alpha \tilde r+\\
 \quad\quad ~~~~~~3{{\tilde L}^2}{{\tilde r}^2} + 3{(\alpha \tilde r)^2} - 2\alpha {{\tilde r}^3} + 10{{\tilde L}^2}\alpha  - 12{{\tilde L}^2}\tilde r + 2\alpha {{\tilde r}^2} - 2{{\tilde r}^3}\Bigg]\,.
\end{array}
 \quad
\end{eqnarray}

 For more details on the derivation of the above equations one can see \cite{Harko:2009kj,Chen:2011wb,Yang:2018wye,Perez:2012bx,Karimov:2018whx,Ding:2019sfy}.
It should be emphasised here that the equation  $ d^{2}\tilde{V}_{eff}/d\tilde{r}^{2} = 0 $ is employed as a necessary condition for the existence of innermost stable circular orbits. In addition to this, the sign of  $ d^{2}\tilde{V}_{eff}/d\tilde{r}^{2} $, usually is a criterion to show the stability of the orbits of the black hole.

By employing all these above mentioned conditions, the location of the event horizons, and of the stable circular orbits, for a S-V-T (QMOG) black hole, surrounded by a
quintessence scalar field, or an exotic matter source, are presented in Table~\ref{tab:alpha}, and Table~\ref{tab:Q}, respectively.

In Table~\ref{tab:alpha}, we did investigate the role of the quintessence parameter $ \tilde{\gamma} $ on the black hole properties, when the MOG parameter $\alpha$ is fixed at $ 0$ and $0.25$. Then, in  Table~\ref{tab:Q}, we considered the effects of changes in the parameter $\alpha$, when the parameter $\tilde{\gamma} $ takes the values $0$ and $0.003$.

In addition to the points mentioned above, the location of the observer is also important. Usually, the
observer is considered to be  at the infinite boundary for the asymptotically flat spacetimes. However,
for the model of quintessence black hole, the cosmological horizon does matter.
Physically, the observer should be located in the domain of outer communication, which is
between the event horizon and cosmological horizon, similar to the  de Sitter case. For
convenience, the observer in our paper is set to be near the cosmological horizon.

\begin{figure}
	\centering
\subfigure{
\includegraphics[scale=0.40]{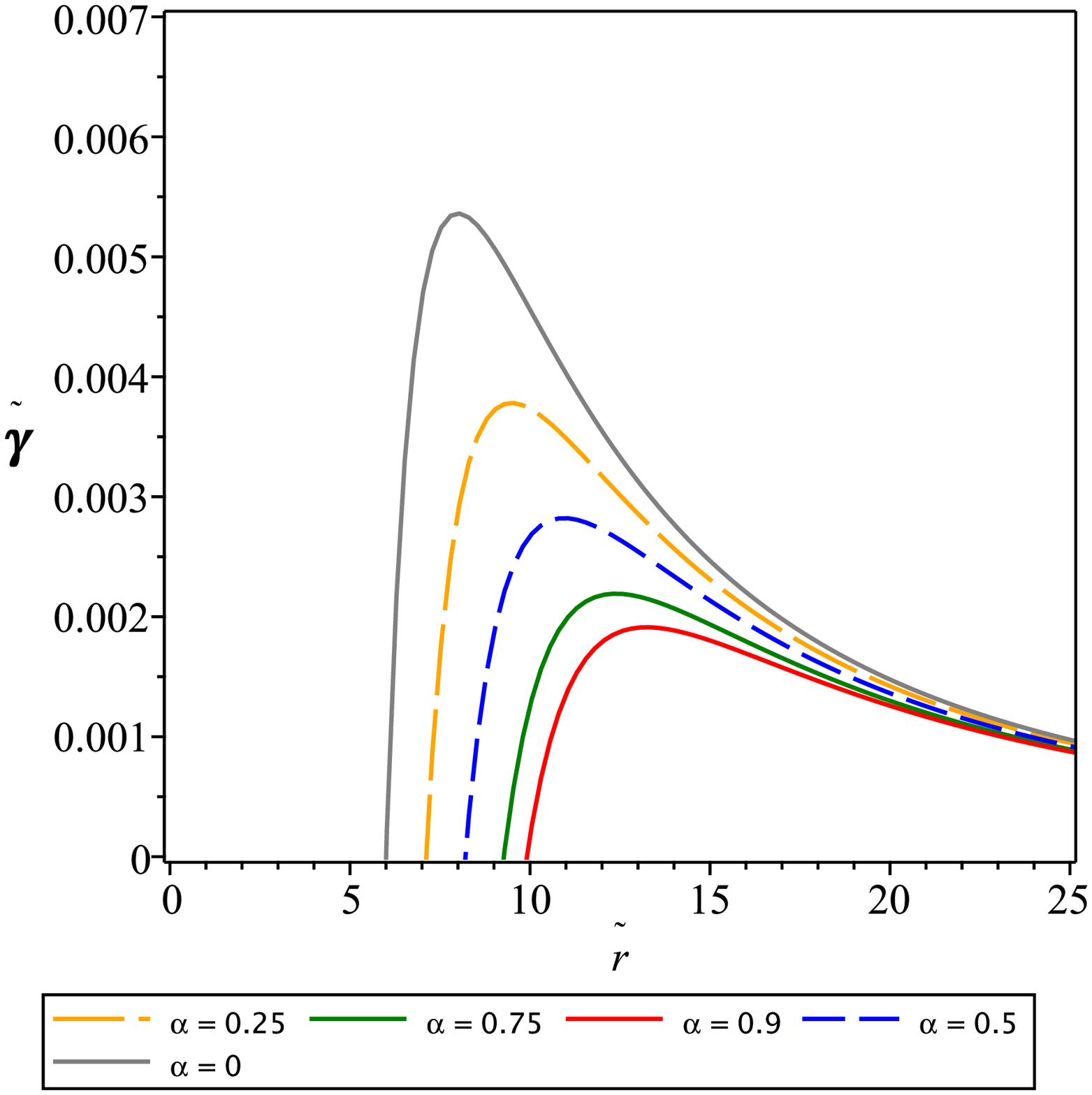}
\includegraphics[scale=0.40]{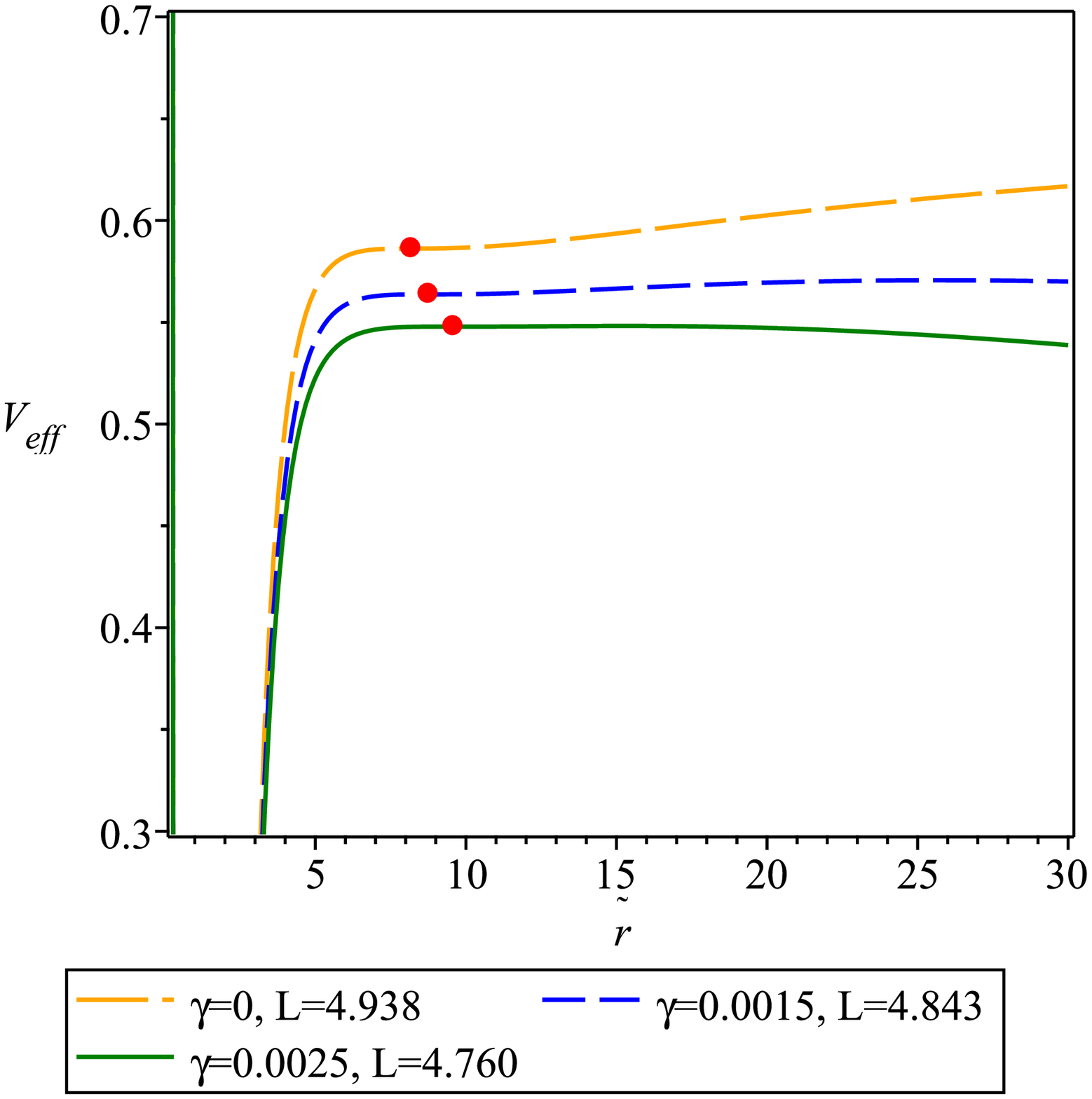}
}
\caption{Right panel: the variation of the quintessence parameter $\tilde{\gamma}$ as a function of the radial coordinate of the event horizon, $\tilde{ r}$, for  different values of the MOG parameter $\alpha$. When $\alpha=0$, the model reduces to the Schwarzschild black hole in the presence of the quintessence field.
Left panel: The variation of the effective potential for the parameter $ \alpha=0.25$, and for different values of
$ \tilde{\gamma} $ and $ \tilde{L} $. The red dots indicate the location of the innermost stable circular orbits, ISCO.}
\label{Pic:Veff}
\end{figure}

\begin{figure}[h]
	\centering
	\subfigure{
		\includegraphics[width=0.5\textwidth]{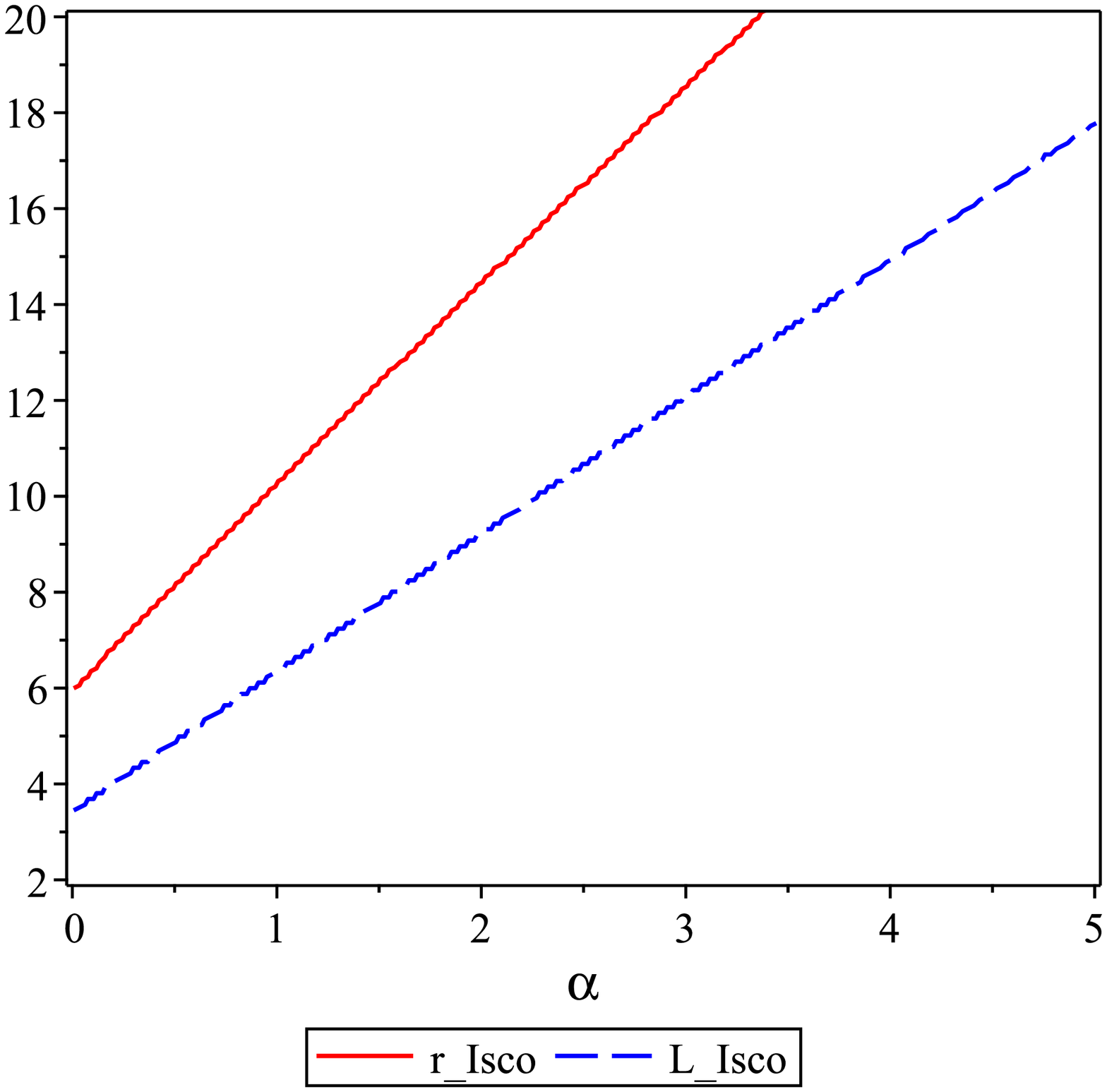}
	\includegraphics[width=0.5\textwidth]{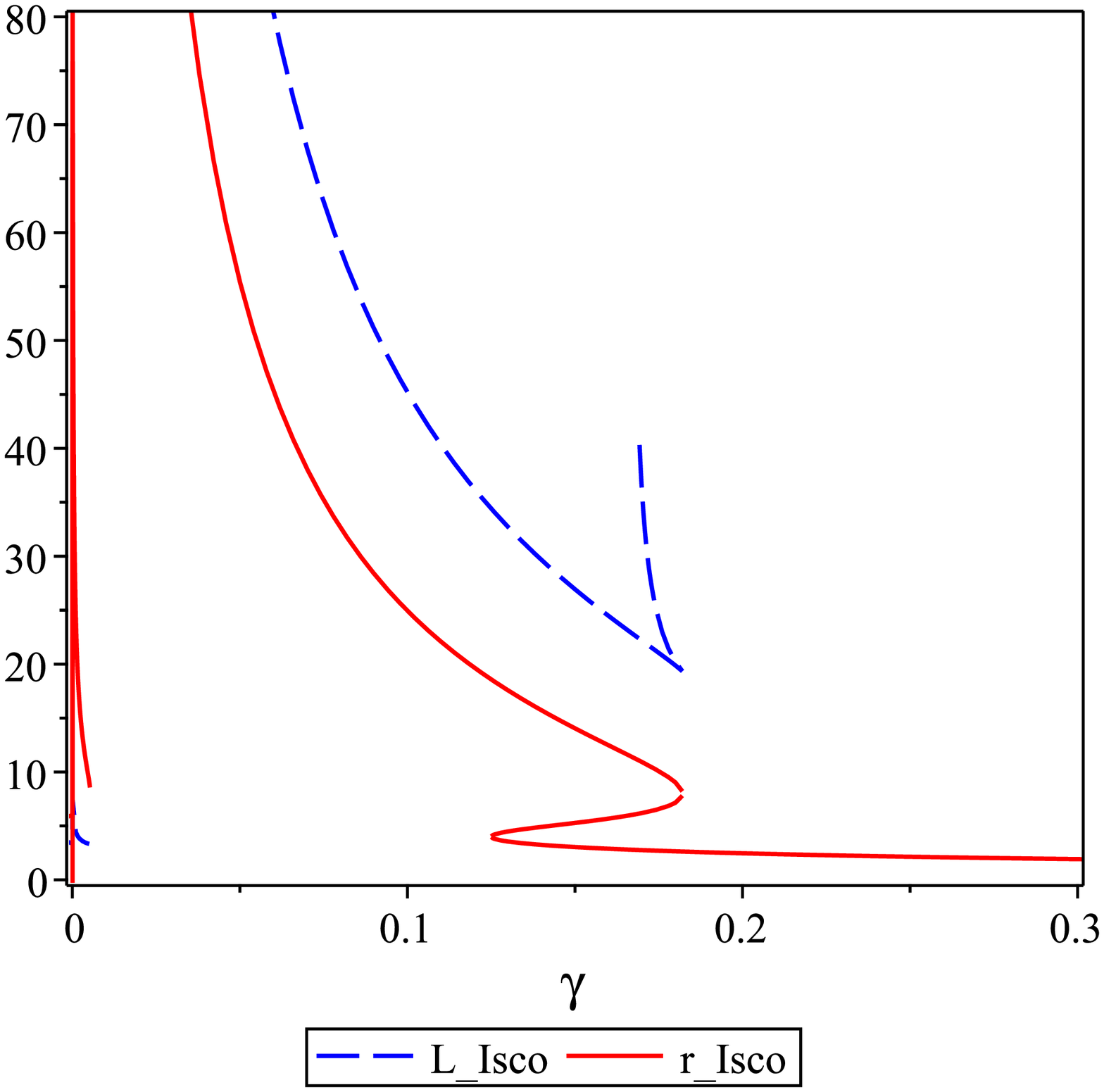}
	}
	\caption{The innermost stable circular orbit (solid line) and the angular momentum of the inflection point (dashed line) in terms of the coupling constant of theory $\alpha$ ($\tilde{\gamma}=0$) (left panel) and $\tilde{\gamma}$ ($\alpha=0$) (right panel).}
	\label{Pic:ISCO12a}
\end{figure}

To visualize the effects of the MOG parameter we present the Fig.~\ref{Pic:Veff}, in which the parameters $\tilde{\alpha}$ and $ \tilde{\gamma} $ are plotted as a function of the radius of the event horizon, for different values of $ \tilde{Q} $ and  $ \tilde{L} $.  It is clear from Fig.~\ref{Pic:Veff} that as long as the values of $ \tilde{Q} $ increase, the peak values of $ \tilde{\gamma} $ become larger. Moreover, considering Fig.~\ref{Pic:Veff}, it can be seen that the increase in the values of $ \tilde{\gamma} $ leads to a decrease of the effective potential. The location of the innermost stable circular orbits is represented, for different values of the parameters of the black hole solution, in Fig.~\ref{Pic:ISCO12a}.


\begin{table}
\begin{tabular}{|p{0.6in}|p{0.5in}|p{0.5in}|p{0.5in}|p{0.5in}|p{0.5in}|p{0.5in}|p{0.5in}|} \hline
& $\gamma$ & $r_-$ & $ r_+$ & $r_c $& $r_{ISCO}$&$ r_S $& $r_{OSCO}$  \\ \cline{2-8}
$ \alpha=0 $	& 0 & $ \times $ & 2 & $ \times $ & 6 & $ \times $ & $ \times $ \\ \hline
	& 0 & 0.1320 & 2.368 & $ \times $ & 7.1106 & $ \times $ & $ \times $ \\ \cline{2-8}
$ \alpha=0.25 $	& 0.001 & 0.1319 &2.3755 & 797.49 & 7.3186 & 24.29 & 2394.7 \\  \cline{2-8}
	& 0.003 & 0.1319 & 2.3907 & 264.14 & 8.0783 & 12.542 & 794.6 \\ \cline{2-8}
		& 0.004 & 0.1319& 2.3984 & 197.469 & 493.59 & $ \times $ & $ \times $\\ \hline
\end{tabular}
\caption{Location of the event horizons, and of the stable circular orbits in a S-V-T (MOG) black hole for $ \alpha=0,~0.25 $, when the quintessence parameter $ \tilde{\gamma} $ takes different values, including $\tilde{\gamma}=0$ to $\tilde{\gamma}=0.004$. Here for the left side of the table one notice $r_-$ stands for Cauchy radius, $r_+$ refers event horizon and $r_c$ is cosmological horizon. Also considering the right panel $r_{ISCO}$ stands for ISCO radius, $r_S$ refers stable radii and $r_{OSCO}$  refers the most outer stable circular  orbit.  }\label{tab:alpha}
\end{table}

\begin{table}
\begin{tabular}{|p{0.5in}|p{0.5in}|p{0.5in}|p{0.5in}|p{0.5in}|p{0.5in}|p{0.5in}|p{0.5in}|} \hline
& $\alpha$ & $r_-$ & $ r_+$ & $r_c $& $r_{ISCO}$&$ r_S $& $r_{OSCO}$  \\ \cline{2-8}
$ \tilde{\gamma} $=0& 0 & $ \times $ & 2 & $ \times $ & 6 & $ \times $ & $ \times $ \\ \hline
	& 0 & $ \times $ & 2.012 & 331.3 & 6.451 & 13.34 & 995.7 \\ \cline{2-8}
    & 0.25 &0.1319 &2.3907 & 264.14 & 8.078 & 12.542 & 794.68 \\ \cline{2-8}
	$ \tilde{\gamma} $=0.003	& 0.5 & 0.1340 & 1.877 & 331.3 & 5.961 & 13.47 & 995.8 \\ \cline{2-8}
	& 0.5 & 0.2751& 2.7628 & 219.18 & 660.27 & $ \times $ & $ \times $\\ \cline{2-8}
	& 0.75 & 0.4269 & 3.1325 & 186.91 & 563.952 & $\times $ & $ \times $\\\hline
\end{tabular}
\caption{The location of the event horizons, and of the stable circular orbits in a S-V-T (MOG) black hole for $ \tilde{\gamma} =0,~0.003 $, and for different values of the MOG parameter $ \alpha $, indicated in the Table.}\label{tab:Q}
\end{table}


 As one can see from Fig.~\ref{Pic:Veff}, the maximum value of the parameter $\tilde{\gamma}$ corresponds to $\tilde{r}\simeq8$. By increasing the value of $\alpha$, the maximum value of the  parameter  $\tilde{\gamma}$ decreases, and the effects of the maximum values of the scalar field do appear far from the MOG black holes.

\section{Astrophysical properties of the QMOG black holes-deflection of light, shadow and the Shapiro delay}\label{sectastro}

In the present Section, for the sake of completeness, we will investigate in detail some of the basic astrophysical tests for the QMOG black hole. In particular, we will consider the deflection of light, the shadow of the black hole, and the Shapiro time delay effect.

\subsection{The deflection of light}

The deflection angle of a photon as it moves from infinity to $r_m$ and off to infinity for the metric \eqref{metric} can be expressed as \cite{19}-\cite{21},
\begin{equation}\label{eq37}
\delta \varphi=\int_{r_{m}}^{\infty}{\dfrac{2dr}{\sqrt{\dfrac{r^{4}}{b^{2}}-f (r)r^{2}}}}-\bar{\pi}=I-\bar{\pi },
\end{equation}
where $ b=\sqrt{r_{m}^{2}/f(r_{m})} $ is the impact parameter of the null ray, and $ r_{m} $ is the coordinate distance of the closest approach. Here $\bar{\pi}$ is the change in the angle $ \varphi $ for the straight line motion, and is therefore subtracted out from the total deflection angle.

We now calculate the integral in Eq.~(\ref{eq37}). Writing the term in the denominator
of \eqref{eq37} as  $f(r) r^2 \left(r^2/b^2 f(r) -1\right)$, for $M \ll r$ one has
\begin{equation}
\begin{aligned}
& \frac{f\left(r_m\right)}{f(r)} \frac{r^2}{r_m^2}-1=\left[\frac{1-\frac{2G M}{r_m}+\frac{G Q^2}{r_m^2}-\frac{G\gamma}{r_m^{3 \epsilon+1}}}{1-\frac{2G M}{r}+\frac{GQ^2}{r^2}-\frac{G\gamma}{r^{3 \epsilon+1}}}\right]\left(\frac{r^2}{r_m^2}\right)-1\, \\
&= \left(\frac{r}{r_m}\right)^2\left[1+2 GM\left(\frac{1}{r}-\frac{1}{r_m}\right)-GQ^2\left(\frac{1}{r^2}-\frac{1}{r_m^2}\right)+G \gamma\left(\frac{1}{r^{3 \epsilon+1}}-\frac{1}{r_m^{3 \epsilon+1}}\right)\right]-1\, \\
& =\left(\frac{r^2}{r_m^2}-1\right)\left[1-\frac{2 GM r}{r_m\left(r+r_m\right)}+\frac{GQ^2}{r_m^2}+\frac{G\gamma r^2}{r^2-r_m^2}\left(\frac{1}{r^{3 \epsilon+1}-r_m^{3 \epsilon+1}}\right)\right]\,.
\end{aligned}
\end{equation}

Upon expanding in powers of $M/r$, $M/r_m$, and $G \gamma$, the integrand in Eq. \eqref{eq37} then becomes,
\begin{equation}\label{eq42}\nonumber
\begin{aligned}
&\int_{r_{m}}^{\infty}\dfrac{2dr}{\sqrt{\dfrac{r^{4}}{b^{2}}-f (r)r^{2}}}=\int\limits_{{r_m}}^\infty {\frac{1}{{\sqrt {\left( {\frac{1}{{r_m^2}} - \frac{1}{{{r^2}}}} \right)} }}} \Bigg\{1 + \frac{{GM}}{r}\left[ {1 + \frac{{{r^2}}}{{{r_m}\left( {r + {r_m}} \right)}}} \right] - \frac{{G{Q^2}}}{2}\left( {\frac{1}{{{r^2}}} + \frac{1}{{r_m^2}}} \right) \\\nonumber
&+\frac{{G\gamma }}{2}\left[ {\frac{1}{{{r^{3 + 1}}}} - \frac{{{r^2}}}{{2\left( {{r^2} - r_m^2} \right)}}\left( {\frac{1}{{{r^{3 + 1}}}} - \frac{1}{{r_m^{3 + 1}}}} \right)} \right]\Bigg\}\frac{{dr}}{{{r^2}}}\,.
\end{aligned}
\end{equation}

After making the substitution $\sin(\theta)= r_{m}/r$, the integral can be expressed as follows,
\begin{equation}\label{eq37aa}
\begin{aligned}
\int_{r_{m}}^{\infty}\dfrac{2dr}{\sqrt{\dfrac{r^{4}}{b^{2}}-f (r)r^{2}}}=\int_0^{\frac{\pi}{2}} d \theta\Bigg\{1+\frac{G M}{r_m}\Bigg(\sin (\theta)+\frac{1}{1+\sin (\theta)}\Bigg)
-\frac{GQ^2}{2 r_m^2}\left(1+\sin ^2(\theta)\right)\\
+\frac{G\gamma}{2 r_m^{3 \epsilon+1}}\left[\sin ^{3 \epsilon+1}(\theta)
 -\frac{1}{2 \cos ^2(\theta)}\big\{\sin ^{3 \epsilon+1}(\theta)-1\big\}\right]\Bigg\} \\
 = \frac{\pi}{2}+\frac{2 G M}{r_m}-\frac{3 \pi GQ^2}{8 r_m^2}+\frac{G \gamma}{4 r_m^{3 \epsilon+1}} \int_0^{\frac{\pi}{2}} \frac{\cos (2 \theta) \sin ^{3 \epsilon+1}(\theta)+1}{\cos ^2(\theta)} d \theta\,.
\end{aligned}
\end{equation}

Gathering together all the results above, the deflection angle is obtained in the form,
\begin{equation}\label{deflect}
\Delta \varphi=\dfrac{4GM}{r_{m}}-\dfrac{3\pi GQ^2}{4r_{m}^2}+\dfrac{G\gamma}{2r_{m}^{3\epsilon+1}}\int_{0}^{\frac{\pi}{2}}\dfrac{\cos(2\theta)\sin^{3\epsilon+1}(\theta)+1}{\cos^{2}(\theta)} d\theta\,,
\end{equation}
a relation that is valid for large $r$ ($r\rightarrow \infty$).

\subsection{The shadow of the QMOG black hole}

Now we are in the position to consider the shadow of QMOG classes of black holes \cite{Falcke:1999pj,EventHorizonTelescope:2019ggy,Perlick:2021aok}. In fact, to obtain the shadow of the black hole we follow up the null geodesics, on which the motion of the photons take place.

The angular radius of the shadow of the black hole, as seen by an observer located at $ r_{0} $, is defined as \cite{ref1},
\begin{equation}
\sin^{2}(\Gamma)=\dfrac{r_{ph}^{2} f(r_{0})}{r_{0}^{2} f(r_{ph})}\,,
\end{equation}
where for small values of $\Gamma$, $\sin(\Gamma)\approx \Gamma$, $ r_{ph} $ denotes the radius of the photon sphere, and $ \Gamma $ is the angle subtended by the radius of the shadow, as seen by a typical observer located at $r_{0}$.

By using the expression of $ f(r) $, as introduced in the metric \eqref{metric}, one gets,
\begin{equation}
\begin{aligned}
\dfrac{r_{ph}^{2}f(r_{0})}{r_{0}^{2}f(r_{ph})} &=\dfrac{r_{ph}^{2}}{r_{0}^{2}}\left[\dfrac{1-\dfrac{2G M}{r_{0}}+\dfrac{GQ^2}{r_{0}^2}-\dfrac{G\gamma}{r_{0}^{3\epsilon+1}}}{1-\dfrac{2G M}{r_{ph}}+\dfrac{GQ^2}{r_{ph}^2}-\dfrac{G\gamma}{r_{ph}^{3\epsilon+1}}} \right]
\nonumber\\
&= \dfrac{r_{ph}^{2}}{r_{0}^{2}}\left[\left( 1-\dfrac{2GM}{r_{0}}+\dfrac{GQ^2}{r_{0}^2}-\dfrac{G\gamma}{r_{0}^{3\epsilon+1}}\right)\left(1
+ \dfrac{2GM}{r_{ph}}-\dfrac{GQ^2}{r_{ph}^2}+\dfrac{G\gamma}{r_{ph}^{3\epsilon+1}}\right)\right] \\ \nonumber
&= \dfrac{r_{ph}^{2}}{r_{0}^{2}}\left[1+2GM\left(\dfrac{1}{r_{ph}}-\dfrac{1}{r_{0}}\right)-GQ^2\left(\dfrac{1}{r_{ph}^{2}}-\dfrac{1}{r_{0}^2}\right)+G\gamma\left(\dfrac{1}{r_{ph}^{3\epsilon+1}}-\dfrac{1}{r_{0}^{3\epsilon+1}}\right) \right],
\end{aligned}
\end{equation}
yielding in turn
\begin{equation}
\sin(\Gamma)=\dfrac{r_{ph}}{r_{0}}+\dfrac{G M(r_{0}-r_{ph})}{r_{0}^{2}}-GQ^2\dfrac{(r_{0}^2-r_{ph}^2)}{2r_{0}^2r_{ph}^2}+\dfrac{G\gamma}{2}\left(\dfrac{1}{r_{ph}^{3\epsilon+1}}-\dfrac{1}{r_{0}^{3\epsilon+1}}\right)\,,
\end{equation}

To get the shape of shadow, we obtain first the unstable circular orbits of the null geodesics. They are determined by the equations,
\begin{equation}
V_{eff}\vert_{r_{ph}}=V_{eff}^{\prime}\vert_{r_{ph}}=0\,.
\end{equation}

The impact parameters are now related as,
\begin{equation}
\dfrac{f(r_{ph})}{r_{ph}^{2}}\left(\kappa+L^{2}\right)-E^{2}=0,\;\;\;\;\dfrac{r_{ph}f^{\prime}-2f(r_{ph})}{r_{ph}^{3}}(\kappa+L^{2})=0,
\end{equation}
where $\kappa$ is the Carter constant, with $r_{ph}$ obtained as a solution of the constraint equation,
\begin{equation}
r_{ph}f^{\prime}(r_{ph})-2f(r_{ph})=0.
\end{equation}

Defining the impact parameters $\xi$ and $\eta$, which are functions of the energy $E$ and of the angular momentum $L$, and of the Carter constant $\kappa$, as \cite{Singh:2017vfr}
\begin{equation}
\xi:=\dfrac{L}{E}\,\,\,\,\eta:=\frac{\kappa}{E^2}\,
\end{equation}
and using the conserved quantity $E$ one can plot the shadow of black holes satisfactorily
\begin{figure}[h]
	\centering
	\subfigure{
		\includegraphics[width=0.3\textwidth]{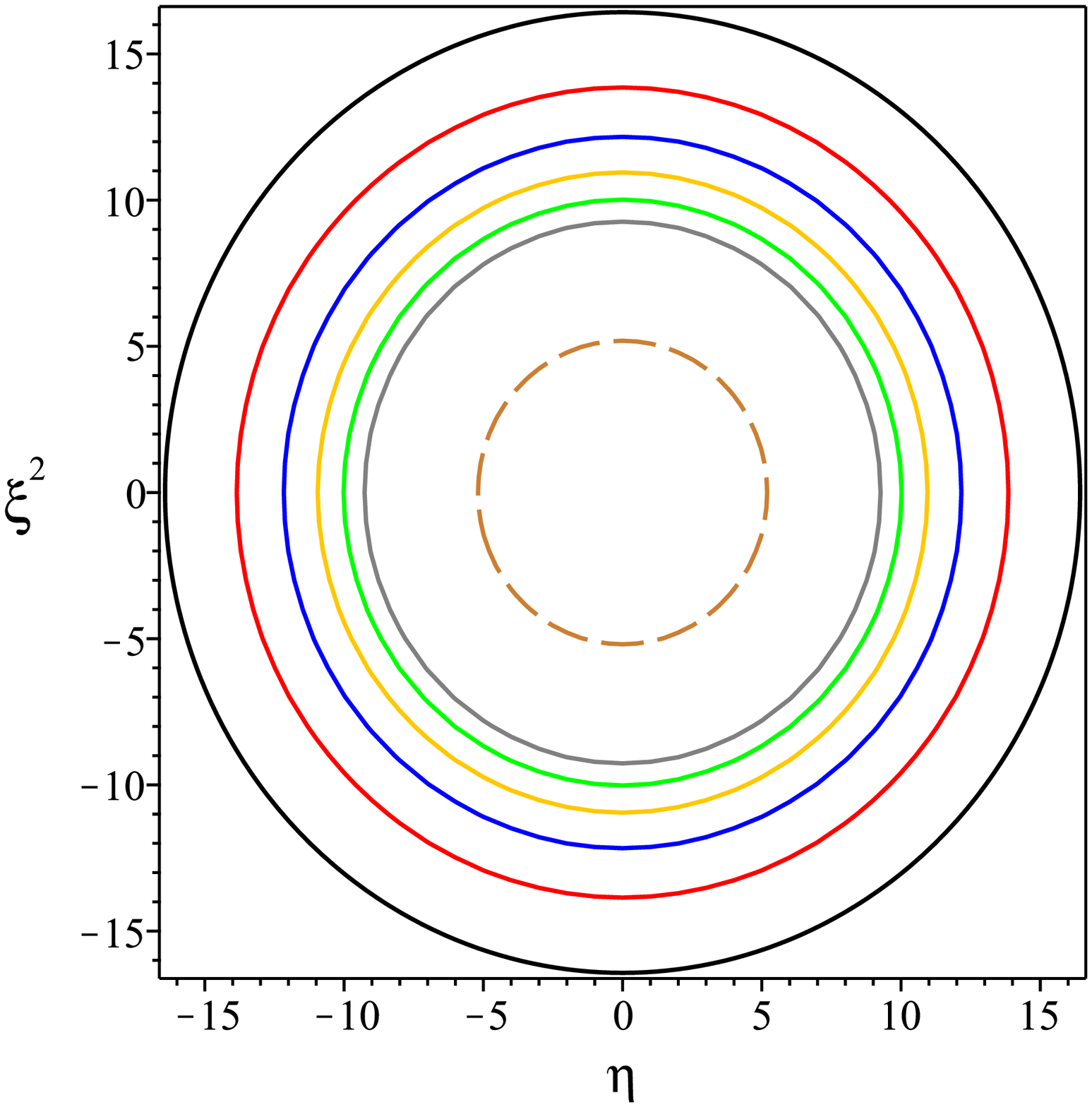}
	\includegraphics[width=0.3\textwidth]{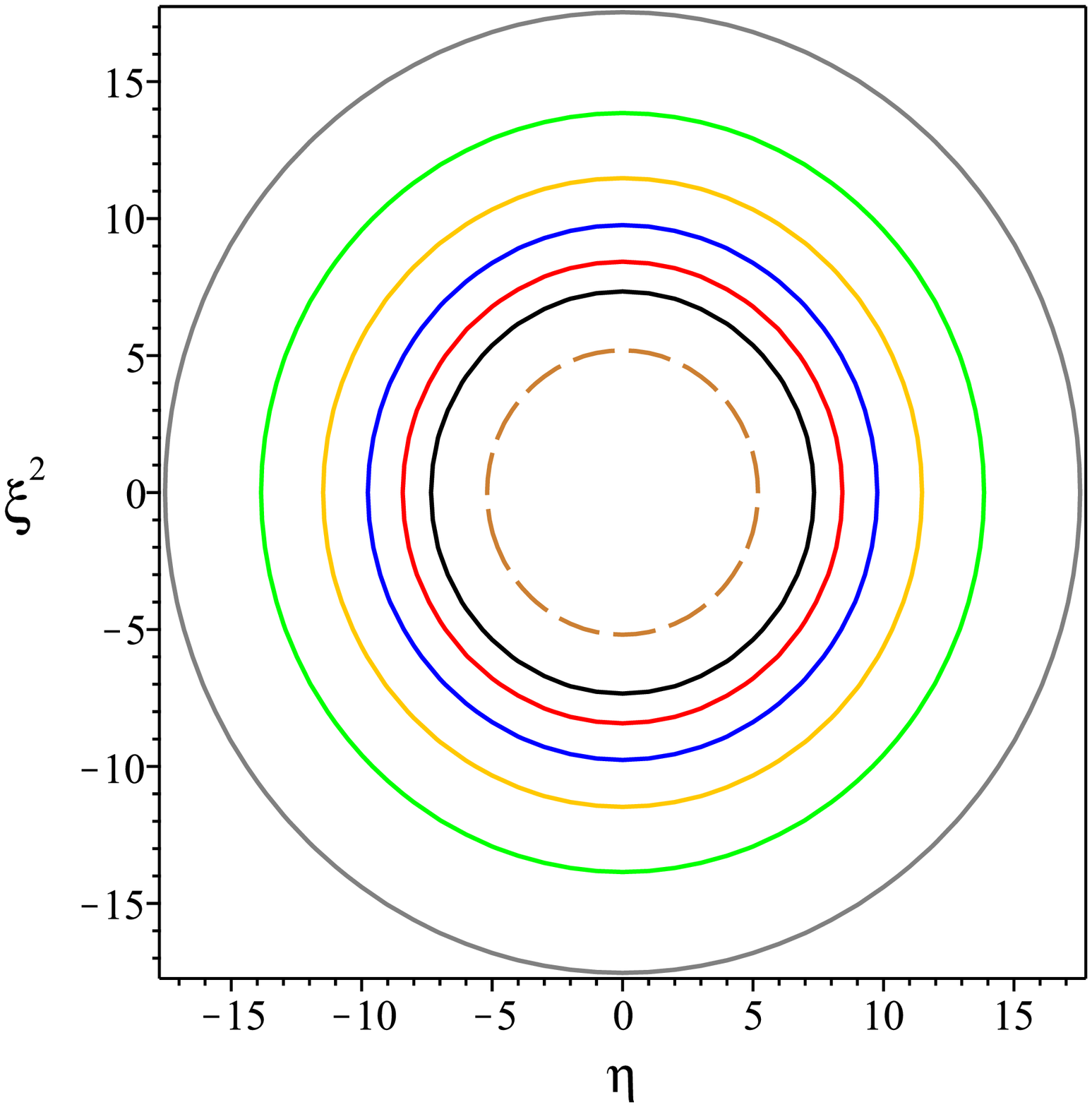}
	}
		\subfigure{
	\includegraphics[width=0.3\textwidth]{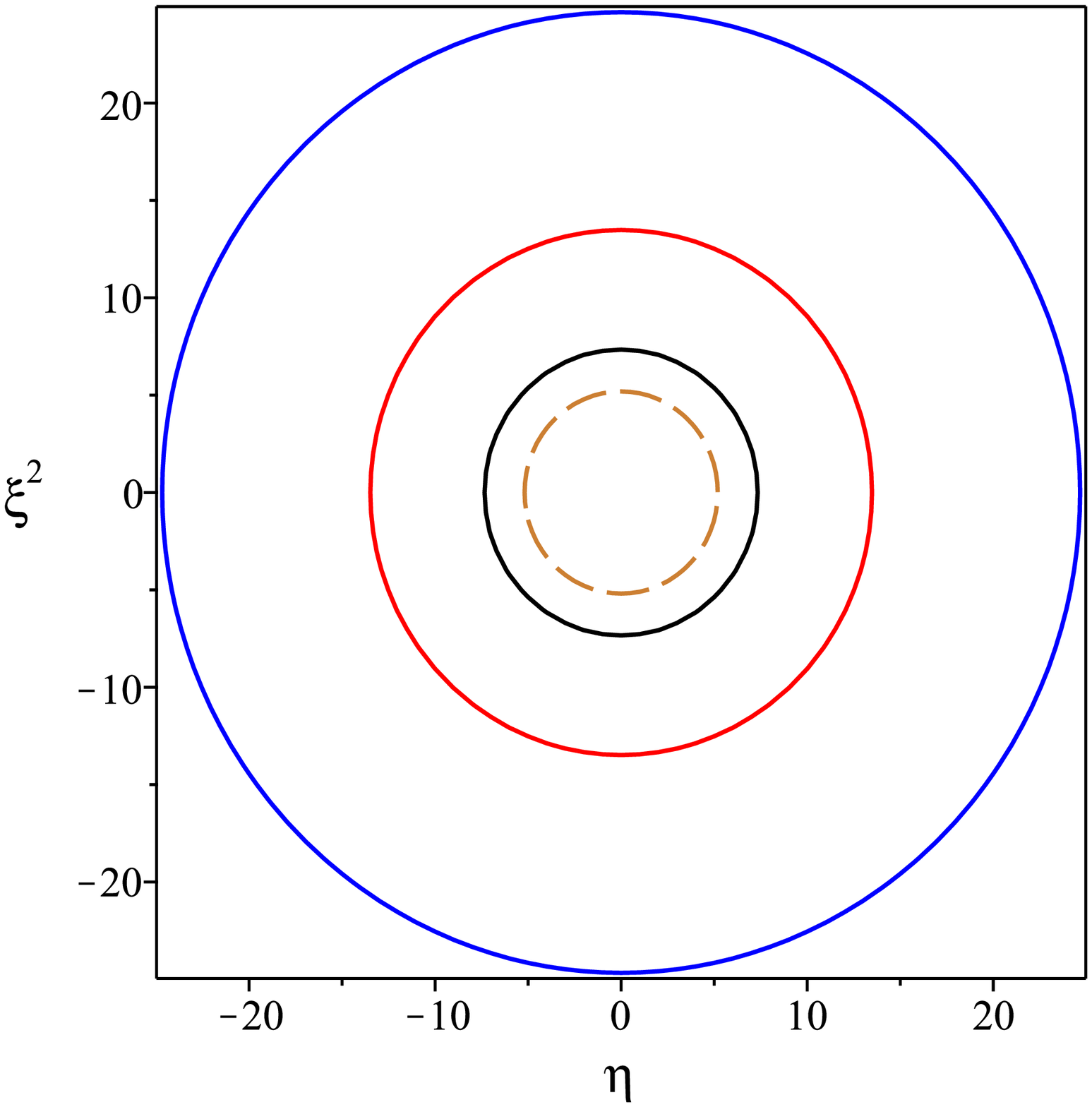}
	}
	\caption{The shadow of the QMOG black hole for $M=1,\epsilon=-2/3,G_{N}=1,\alpha=0.5$ and $\gamma={0.045},\textcolor{red}{0.04},0.035,0.03,0.025$ (left panel), and for $\gamma=0.04$ and $\alpha=0.1,\textcolor{red}{0.2},0.3,0.4,0.5,0.6$ (Middle panel), and for $\alpha=0.1,\gamma=0.04$ and $M=1,\textcolor{red}{1.5},2$ (right panel). Here the dashed line presents the shadow of  the pure Schwarzschild solution with $\alpha=\gamma=0$.
}
	\label{Pic:ISCOShadaow}
\end{figure}

The shape of the shadow of the QMOG black hole is presented in Fig.~\ref{Pic:ISCOShadaow}. Since, the black hole is static, the shadow is circular.
As reported in \cite{EventHorizonTelescope:2021srq}, for the $M87$ galaxy, the angular diameter of the shadow of the central black hole is $\theta_{M87}=42\pm3\mu as$ as, the distance of the $M87$ from the Earth is $D =16.8 Mpc$, and the mass of the $M87$ is $M_{M87} = 6.5\pm0.9\times10^{9} M_{\odot}$. Similarly, for Sagittarius A*, the observational data are provided in the recent EHT paper \cite{EventHorizonTelescope:2022wkp}. The angular diameter of the shadow is $\theta_{Sgr.A*} = 48.7\pm7\mu as$ as (EHT), the distance of the Sgr. A* from the Earth is $D = 8277\pm33 pc$, and mass of the black hole is $M_{Sgr. A*} = 4.3\pm0.013\times10^{6} M_{\odot}$ \cite{EventHorizonTelescope:2022wkp}.

Now, once we have the above data about the black holes, we can calculate the diameter of the shadow size in units of mass, by using the following expression,
\begin{equation}
d_{sh}=\dfrac{D\theta}{M},
\end{equation}

\begin{figure}[h]
	\centering
	\subfigure{
		\includegraphics[width=0.4\textwidth]{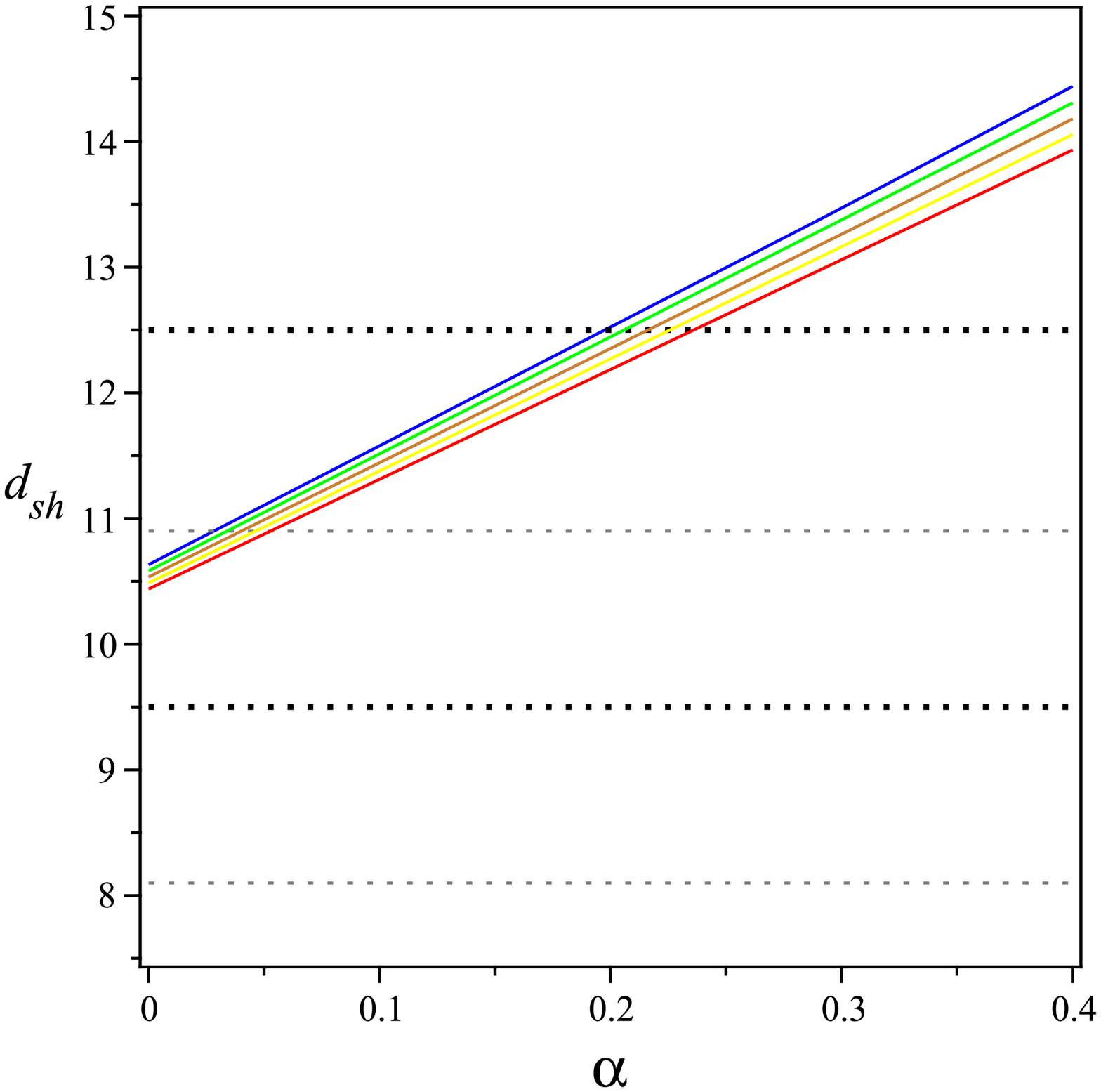}
	\includegraphics[width=0.4\textwidth]{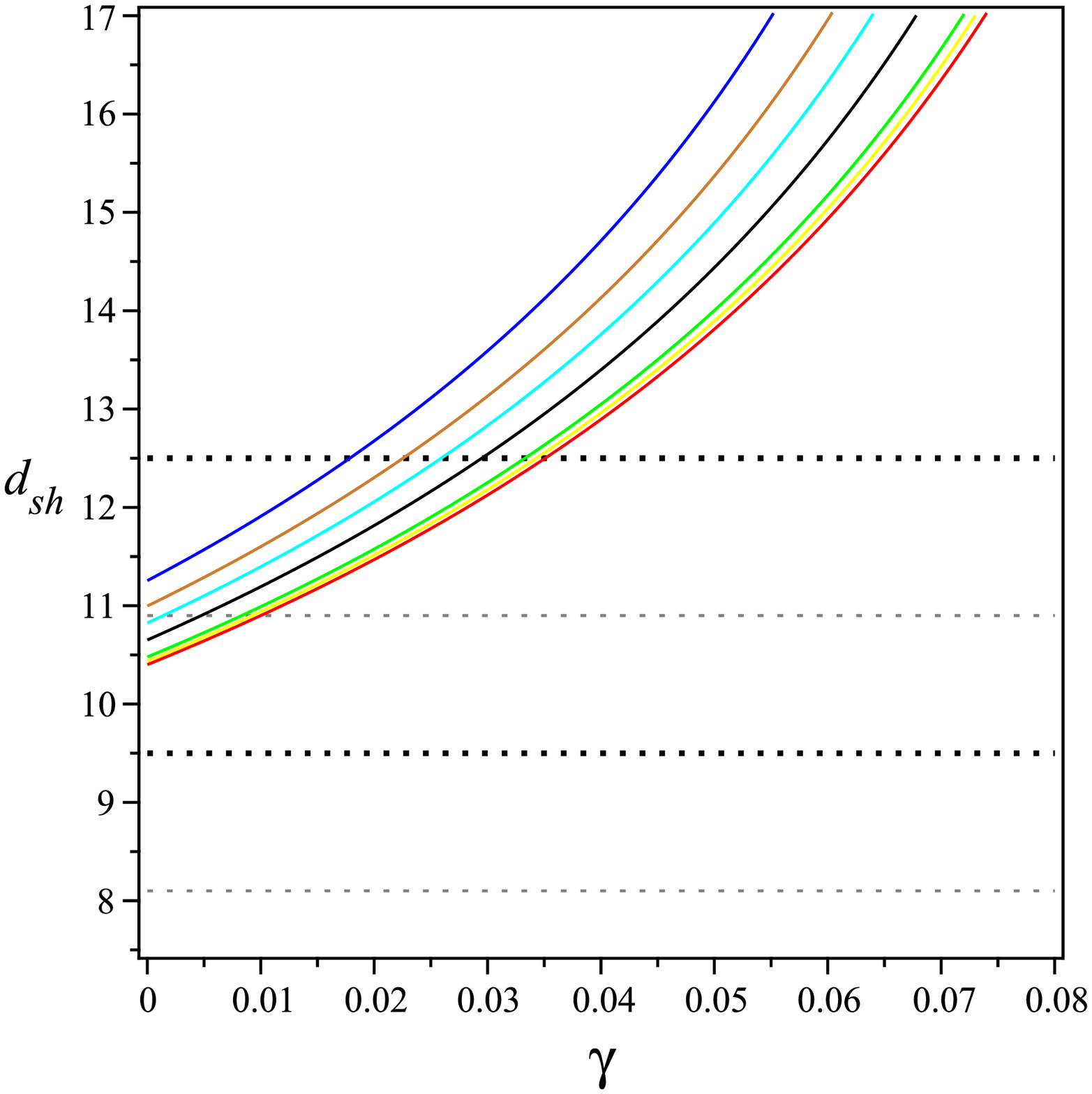}
	}
	\caption{The plots of diameter of the shadow of the black hole in terms of $\alpha$ for $M=1,\epsilon=-2/3,G_{N}=1,\gamma=\textcolor{red}{0.001},0.002,0.003,0.004,0.005$ (left panel) and in terms of $\gamma$ for $\alpha=\textcolor{red}{0.001},0.005,0.01,0.05,0.1$ (right panel).
}
	\label{Pic:ISCOyyy}
\end{figure}

Hence, the theoretical shadow diameter, however, can be obtained via $d^{\theta}_{sh} = 2R_{sh}$. Therefore, by using the above expression, we obtain the diameter of the shadow image of $M87$ $d^{M87}_{sh} = (11\pm1.5)M$, and for Sgr. A* $d^{Sgr.A*}_{sh} = (9.5\pm1.4)M$. The variation of the diameter of the shadow image with coupling parameter $\alpha$ and $\gamma$ for $M87$ and for Sgr. A is shown in Fig.~ \ref{Pic:ISCOyyy}. We observe that there is a lower bounds for $\alpha$ and $\gamma$.

\subsection{The Shapiro time delay}

Finally, we consider the Shapiro time-delay to obtain a bound on the coupling constant $\alpha$. The general expression for the time delay for a photon moving in the metric of the QMOG black hole reads \cite{Shapiro:1964uw},
\begin{equation}
t(r_{0},r)=\int_{r_{0}}^{r}{\dfrac{dr}{f(r)\sqrt{1-\dfrac{r_{0}^{2}}{r^{2}}\dfrac{f(r)}{f(r_{0})}}}}
\end{equation}

In order to evaluate the integral, we expand the metric in the asymptotic regime. Similar manipulations as before yield for the integral,
\begin{equation}
\begin{aligned}
\int_{r_{0}}^{r}{\dfrac{dr}{f(r)\sqrt{1-\dfrac{r_{0}^{2}}{r^{2}}\dfrac{f(r)}{f(r_{0})}}}}=\frac{1}{\sqrt{1-\frac{r_{0}^{2}}{r^{2}}}}\Bigg\{&1+\frac{G M}{r}\left[ 1+\dfrac{r_{0}}{(r+r_{0})}\right]-\frac{3GQ^2}{2r^2}\\
&+G\gamma\left[\frac{r_{0}^2}{2(r^2-r_{0}^2)}\left(\frac{1}{r_{0}^{3\epsilon+1}}-\frac{1}{r^{3\epsilon+1}}\right)+\frac{1}{r^{3\epsilon+1}}\right]\Bigg\}\,.
\end{aligned}
\end{equation}

Working in the asymptotic regime (to 2nd order in the continued fraction expansion), approximately the for the time delay we obtain the expression,
 \begin{equation}
t(r,r_{0})=t_{SR}(r,r_{0})+\Delta t(r,r_{0})
\end{equation}
where $ t_{SR}=\sqrt{r^{2}-r_{0}^{2}} $ is the special relativistic contribution of the propagation of light in
the flat spacetime.

Hence, the maximum round-trip excess time delay is given by,
\begin{equation}
\Delta t(r,r_{0})=2\left[ t(r_{2},r_{0})+t(r_{1},r_{0})-\sqrt{r_{1}^{2}-r_{0}^{2}}-\sqrt{r_{2}^{2}-r_{0}^{2}}\right].
\end{equation}

Here  $ r_{0} $ is the distance of the closest approach of the radar wave to the center of the Sun, $ r_{1} $ is the distance along the line of sight from the Earth to the point of closest approach to the Sun, and $ r_{2} $ represents the distance along the path from this point to the planet, where $ r_{1,2}\gg r_{0}$.
In the case of $ r_{1}=r_{2}=r $,  this becomes,
\begin{equation}
\begin{aligned}
\Delta t(r,r_{0})&=4G M\ln\left(\dfrac{r+\sqrt{r^2-r_0^2}}{r_0}\right)\\
&+4GM\sqrt{\dfrac{r-r_{0}}{r+r_{0}}}-\dfrac{3\pi GQ^2}{r_{0}}\left[1-\dfrac{2}{\pi}\tan^{-1}\left(\dfrac{r_{0}}{\sqrt{r^2-r_{0}^2}}\right)\right]\\
&+2G\gamma \int_{r_{0}}^{r}\dfrac{r(r_{0}^{-3\epsilon+1}-3r_{0}^{2}r^{-3\epsilon-1}+2r^{-3\epsilon+1})}{(r^{2}-r_{0}^{2})^{\frac{3}{2}}}dr\,.
\end{aligned}
\end{equation}\label{eqshap}

Here it is supposed that $r_1=r_2\approx10^{11}\; {\rm m}$ is the distance from the Earth to the point of the closest approach to the Sun, and $r_0\approx10^8\; {\rm m}$  is the distance of closest approach of the signal to the center of the Sun. After integration, and some manipulations, for $Q=0$, and $\epsilon=-2/3$, one gets,
\begin{equation}\label{ShapiroTime}
\begin{aligned}
\Delta t(r,r_{0})=4G M\ln\left(\dfrac{r+\sqrt{r^2-r_0^2}}{r_0}\right)+4GM\sqrt{\dfrac{r-r_{0}}{r+r_{0}}}
+2\gamma G\,{\frac { \left( {{ r_0}}^{2}+r{ r_0}+{r}^{2} \right)  \sqrt{\left( r-
{r_0} \right) }}{\sqrt {{{r}}+{{ r_0}}}}}.
\end{aligned}
\end{equation}

The deviation of the Shapiro time delay from the prediction of the Einstein's general relativity is smaller than $1.2\times 10^{-5}\;{\rm s}$  \cite{Edery:1997hu,Sajadi:2020axg}. By using the Solar System data in SI units, one can constrain the parameter of the QMOG black hole model as $\gamma < 10^{-17}$. To obtain this upper bound, we have assumed that the parameter $\alpha$ is much less than unity in the Solar System tests. This bound was predictable, since in the local regions the effects of the quintessence scalar field is usually very small, as compared to the cosmological scales.

\section{Electromagnetic emissivity of the thin accretion disks around QMOG black holes}\label{Fluxetc...}

The first systematic Newtonian theory of the thin accretion disks existing around black holes dates back to the seminal work by Shakura and Sunyaev in 1973 \cite{Shakura:1972te}. In their paper, they showed that by considering the effects of the friction between different layers of the Keplerian orbits around the black hole, the motion around inward spirals resulted in the loss of angular momentum.  They concluded that due to such a mechanism, the kinetic energy of the disk increases. Moreover,  the release of the gravitational energy also takes place. The warming up of the disk leads to the possibility of the emission of a significant fraction of electromagnetic energy.

The generalization to the general relativistic case has been done by Novikov, Thorne, and Page and Thorne, in 1973 and 1974, \cite{NovikovThorne1973} and \cite{Page:1974he}, respectively. Therefore,  more realistic astrophysical disk models can be constructed in this framework. For their analysis they considered a stationary black hole, which obeys axially or spherical symmetry, like, for example, the Schwarzschild black hole. In the Novikov-Thorne model one considers  that the central plane of the disk coincides with the equatorial plane of the black hole. This assumption significantly decreases the mathematical complexity of the analysis, and leads to the result that all the diagonal and off-diagonal components of the metric depend only on the radial coordinate $r$ \cite{Page:1974he}. Recently, in \cite{Sheikhahmadi:2021uwo,Sheikhahmadi:2020snl} it was shown that for a Schwarzschild  background metric, for a specific type of accretion disk that contains force free magnetic field \cite{Blandford:1977ds}, the form of the perturbed metric resulted in some  modifications not only in diagonal components of the background metric, but also in the appearance of some off-diagonal components, with all of them showing a radial coordinate dependence behaviour. One may consider these results as a confirmation for the assumptions made by the authors of \cite{Page:1974he}.

By taking into account all the points discussed  above, we are now in a position to study the electromagnetic emissivity properties, and the physical properties for the QMOG black hole, and of its proposed disk. The flux $ F $ of the radiant energy over the disk can be written as \cite{NovikovThorne1973,Harko:2009kj,Chen:2011wb,Yang:2018wye,Perez:2012bx,Avara00,Karimov:2018whx,Ding:2019sfy}
\begin{equation}\label{F}
F(\tilde{r})=\dfrac{-\dot{M}_{0}}{4 \pi \sqrt{-g}}\dfrac{\Omega_{,\tilde{r}}}{(\tilde{E}-\Omega \tilde{L} )^{2}}\int^{r}_{r_{\tiny Isco}}(\tilde{E}-\Omega \tilde{L} )  \tilde{L}_{,\tilde{r}} d\tilde{r}\,,
\end{equation}
where $\dot{M}_{0}$ denotes the time averaged  value of the accreted rest mass, which has no dependency to the radial coordinate, and is taken as a constant \cite{Page:1974he}. Substituting Eqs. \eqref{E}, \eqref{L} and \eqref{Om} into \eqref{F} one can obtain for the radiation flux the expression
\begin{eqnarray}\label{Flux-final}
\begin{array}{l}
F(\tilde{r})= - \frac{{8(\alpha  + 1)\dot{M}_{0}\left( {\alpha  - \frac{{3\tilde r}}{4} + \frac{{\tilde \gamma {\tilde{r}^3}}}{8}} \right)}}{{\pi {\tilde{r}^4}\left( {2\alpha  - 2\tilde r + \tilde \gamma {{\tilde r}^3}} \right){{\left( {4{\alpha ^2} + 4\alpha  - 6(\alpha  - 1)\tilde r + 2{{\tilde r}^2} + \tilde \gamma ( - \alpha  - 1){{\tilde r}^3}} \right)}^2}}} \times \\
 \\
\left[ \begin{array}{l}
{\alpha ^2}(\alpha  + 1) - \frac{  9}{4}\left( {{\alpha ^2} +\alpha } \right)\tilde r + \frac{3}{2}\left( {\alpha  + 1} \right){{\tilde r}^2} + \frac{1}{8}\left( {11{\alpha ^2}\tilde \gamma  + 11\alpha \tilde \gamma  - 2} \right){{\tilde r}^3}-\\
 \frac{3}{2}(\alpha  + 1)\tilde \gamma {{\tilde r}^4} + \frac{{3 }}{8}{\tilde \gamma{\tilde r}^5} - \frac{1}{8}(\alpha  + 1){{\tilde \gamma }^2}{{\tilde r}^6}
\end{array} \right]\,.
\end{array}
\end{eqnarray}

\begin{figure}[h]
	\centering
	\subfigure{
		\includegraphics[width=0.5\textwidth]{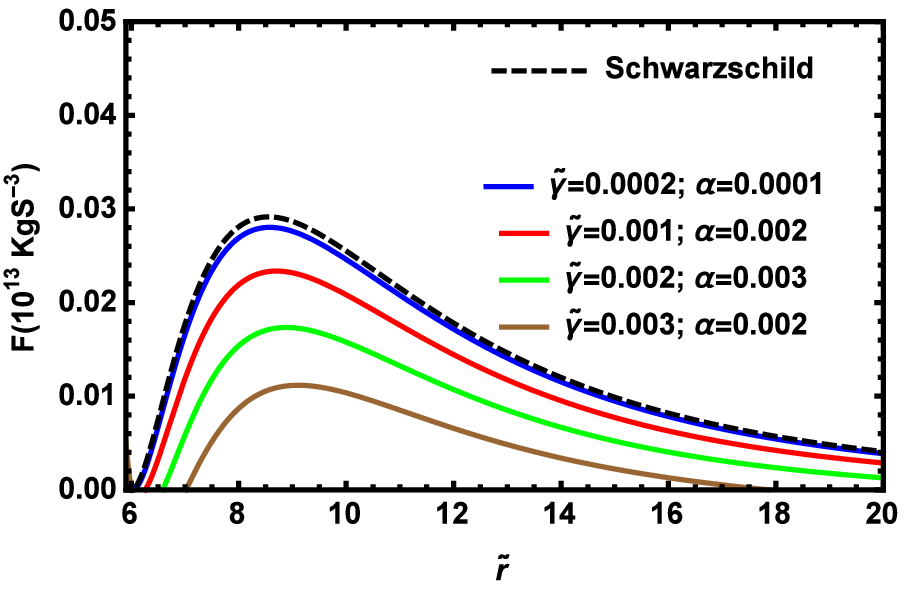}
	\includegraphics[width=0.5\textwidth]{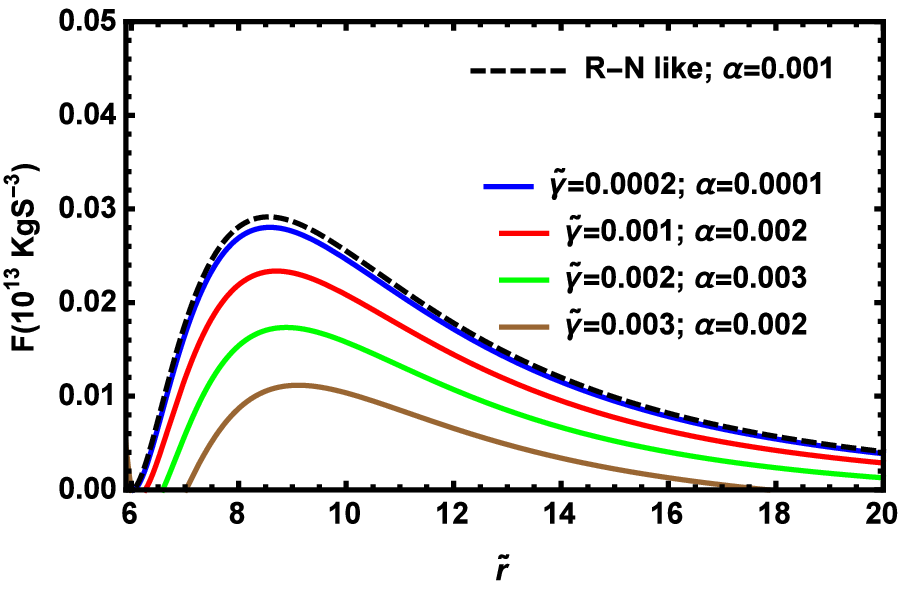}
	}
	\caption{The electromagnetic energy flux versus $\tilde{r}$, for different values of the free parameters of the QMOG black hole model $\alpha$ and $\tilde{\gamma}$. In the left panel  the Schwarzschild case is also presented. In the right panel a comparison with the Reissner-Nordstrom type black hole, with $\gamma =0$, is depicted. In all cases we have considered $\dot{M}_{0}=1.5\times 10^{17} kgs^{-1}$.}
	\label{Pic:Flux-epsilonminus}
\end{figure}

Using the above expression for $F(\tilde{r})$, the variation of the energy flux from the disk is presented in Fig.~ \ref{Pic:Flux-epsilonminus}. In these plots different values for the MOG and quintessence parameters are considered. Obviously, when both $\alpha$ and $\tilde \gamma$ tends to zero, the configuration goes back to the usual Schwarzschild solution. As an another interesting choice, one can let $\tilde \gamma=0$, but  $\alpha$ not, which we call it a Reissner-Nordstr\"{o}m like solution. From these plots, it is immediately seen that by increasing the values of these parameters the peak of the flux, of the is decreasing.

\begin{figure}[h]
	\centering
	\subfigure{
		\includegraphics[width=0.5\textwidth]{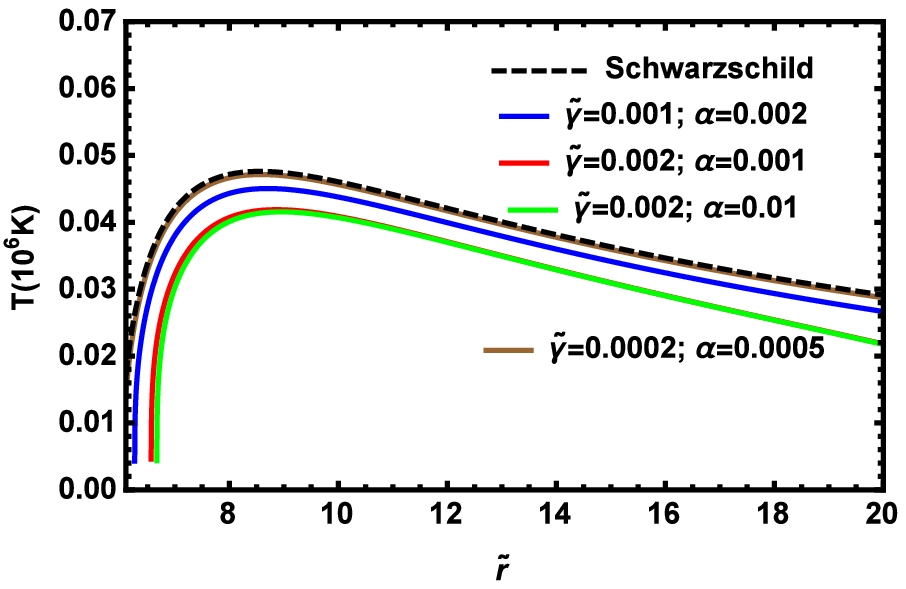}
	\includegraphics[width=0.5\textwidth]{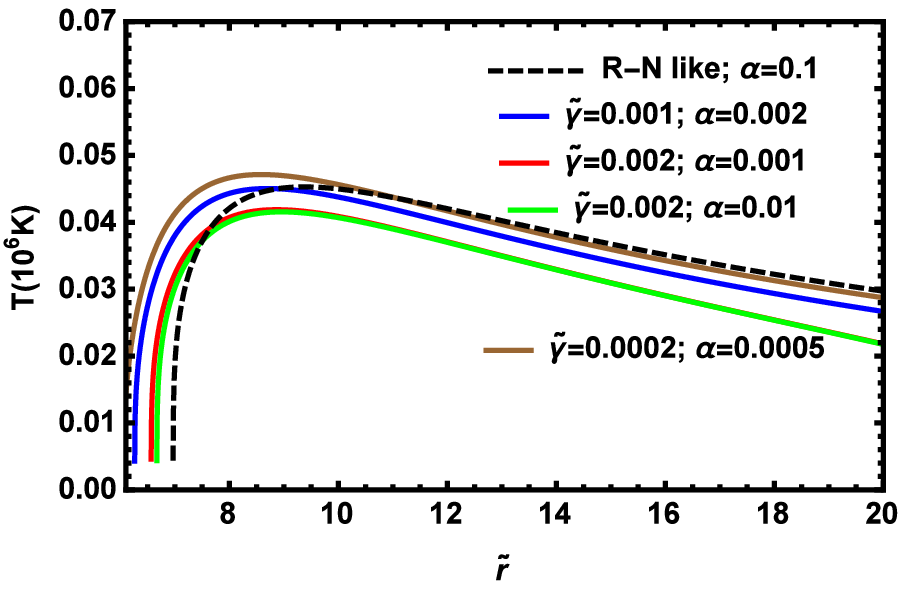}
	}
	\caption{The behaviour of the disk temperature  against $\tilde{r}$, for different values of $\alpha$ and $\tilde{\gamma}$. In the two panels the results based on the Schwarzschild and Reissner-Nordstrom type black holes are also presented. For the Stefan-Boltzmann constant we have adopted the value $\sigma=5.67\times10^{-8}~W m^{-2}K^{-4}$.}
	\label{Pic:alphaT}
\end{figure}

If one assumes that the radiation flux emitted by the disk's surface is in thermodynamical equilibrium, from the Stefan–Boltzmann law \cite{Karimov:2018whx,Ding:2019sfy},
\begin{equation}\label{Sigma}
 F(r)=\sigma T^{4}(r)\,,
\end{equation}
where  $ \sigma $ is the Stefan–Boltzmann constant, one can introduce the temperature $T$ of the accretion disk, see Fig.\eqref{Pic:alphaT}.

Another important physical parameter of the accretion disks is
the efficiency parameter $\mathbf{\mathcal{\zeta}}$, which measures the amount of
 the accreting mass transformed into radiation, in the presence of the central compact object  \cite{Avara00,Karimov:2018whx,Ding:2019sfy}. The
efficiency is measured at infinity, and it is defined as the ratio
of two rates: the rate of energy of the photons emitted from
the disk surface, and the rate with which the mass-energy is
transported to the central body \cite{Avara00}.

If all photons reach infinity,
an estimate of the efficiency is given by the specific energy
of the particles in the disk measured at the marginally stable
orbit \cite{Avara00,Karimov:2018whx,Ding:2019sfy}
\begin{equation}\label{Energy00}
\mathbf{\mathcal{\zeta}}=1-\tilde{E}(r_{_{Isco}})\,.
\end{equation}

The behaviour of the efficiency of the disk in the QMOG geometry is provided in Table.\eqref{tabIII}

\begin{table}
\center
\begin{tabular}{|p{0.6in}|p{0.5in}|p{0.5in}|p{0.5in}|p{0.5in}|p{0.5in}|p{0.5in}|p{0.5in}|} \hline
& $\gamma$ & $r_{ISCO}$&$ \zeta$ \\ \cline{2-4}
  $ \alpha=0 $ & 0 & 6 & $ 0.0571$  \\ \hline
	& 0 & 7.1106 & $ 0.15927 $  \\ \cline{2-4}
$ \alpha=0.25 $	& 0.001 &7.3186 & 0.16694  \\  \cline{2-4}
	& 0.003 & 8.0783 & 0.18313 \\ \cline{2-4}
		& 0.004& 493.59 & $ \times $ \\ \hline
\end{tabular}
\caption{By introducing the amount of $r_{Isco}$ provided in Tabel.\eqref{tab:alpha}, the behaviour of efficiency parameter is investigated. }\label{tabIII}
\end{table}

\section{Hawking radiation of the QMOG black holes}

In the present Section  we would like to study the mass loss rate due to the Hawking radiation, that is $\dot{M}$, with the aim of determining the lifetime of the black hole \cite{Page:2004xp}. It is an already known result that if the mass of a black hole is sufficiently large, then the temperature will be low. Accordingly, without affecting the generality, one can suppose that the mass loss phenomenon that appears due to the  Hawking radiation mechanism originates from the emission of massless particles. In a similar approach used to determine the relation between the flux parameter and the temperature, the mass loss rate is given by  Stefan-Boltzmann law as follows \cite{sbr},
\begin{equation}
\dot{M}=-\dfrac{\pi^2}{60}(\sigma_{g}+\sigma_{\gamma}+\dfrac{21}{8}\sigma_{\nu}) T^{4}\,,
\end{equation}
where $\sigma_{i}$,  are the thermally averaged cross sections of the black hole for gravitons, $g$, photons, $\gamma$, and neutrinos, $\nu$. In the following we define,
\begin{equation}
\rho=\left(\sigma_{g}+\sigma_{\gamma}+\dfrac{21}{8}\sigma_{\nu}\right)\dfrac{1}{\sigma_{0}}\,,
\end{equation}
where $\sigma_{0}$ is the geometrical optics cross section of the black hole. The cross sections for neutrinos, photons and gravitons can be estimated as,
\begin{equation}
\sigma_{\nu}\sim 0.67\sigma_{0},\;\;\;\;\sigma_{\gamma}\sim 0.24\sigma_{0},\;\;\;\;\sigma_{g}\sim 0.03\sigma_{0}.
\end{equation}

To obtain the above values we have assumed $\rho\sim 2.02$, and $\sigma_0$ should be determined as follows. Since the emitted particles move along null geodesics, from Eq.~\eqref{dr/ds}, it follows that their motion is governed by the equation,
\begin{equation}
\dot{r}^2=E^2+f(r)\left(\mu-\dfrac{L^2}{r^2}\right)\,,
\end{equation}
where, as previously discussed, $E$ and $L$ refer to the energy, and the angular momentum, respectively.

 For the emitted particles to reach the infinity, rather than to fall back into the black hole horizon, one has to impose the following condition,
\begin{equation}\label{eqq5.5}
\dfrac{1}{l^2}\equiv\dfrac{E^2}{L^2}\geq f\left(\dfrac{1}{r^2}-\dfrac{\mu}{L^2}\right)\,.
\end{equation}
where by $\mu$ we have denoted the mass of the test particles. The geometrical optical cross section of the black hole is given by,
\begin{equation}\label{eqq4.12}
\sigma_{0}=\pi l_{c}^{2}=\dfrac{\pi r_{c}^2}{f(r_{c})}\dfrac{1}{1-\dfrac{\mu r_{c}^2}{L^2}}.
\end{equation}
Using Eq.~\eqref{eqq5.5}, one can obtain an equation for $r_{c}$ as follows,
\begin{equation}
6GMr_{c}-4GQ^{2}+3G\epsilon\gamma r_{c}^{-3\epsilon+1}+3G\gamma r_{c}^{-3\epsilon+1}-2r_{c}^2=0\,.
\end{equation}
For $Q=0$, and $\epsilon=-2/3$, we have,
\begin{equation}
r_{c}=\dfrac{1-\sqrt{1-6G^2\gamma M}}{G\gamma}.
\end{equation}

For small $\gamma$ values, we obtain
\begin{equation}\label{eqq4.14}
r_{c}\sim 3GM+\dfrac{9G^3M^2\gamma}{2}+\mathcal{O}(\gamma^2)
\end{equation}

\begin{figure}[h]
	\centering
	\subfigure{
		\includegraphics[width=0.5\textwidth]{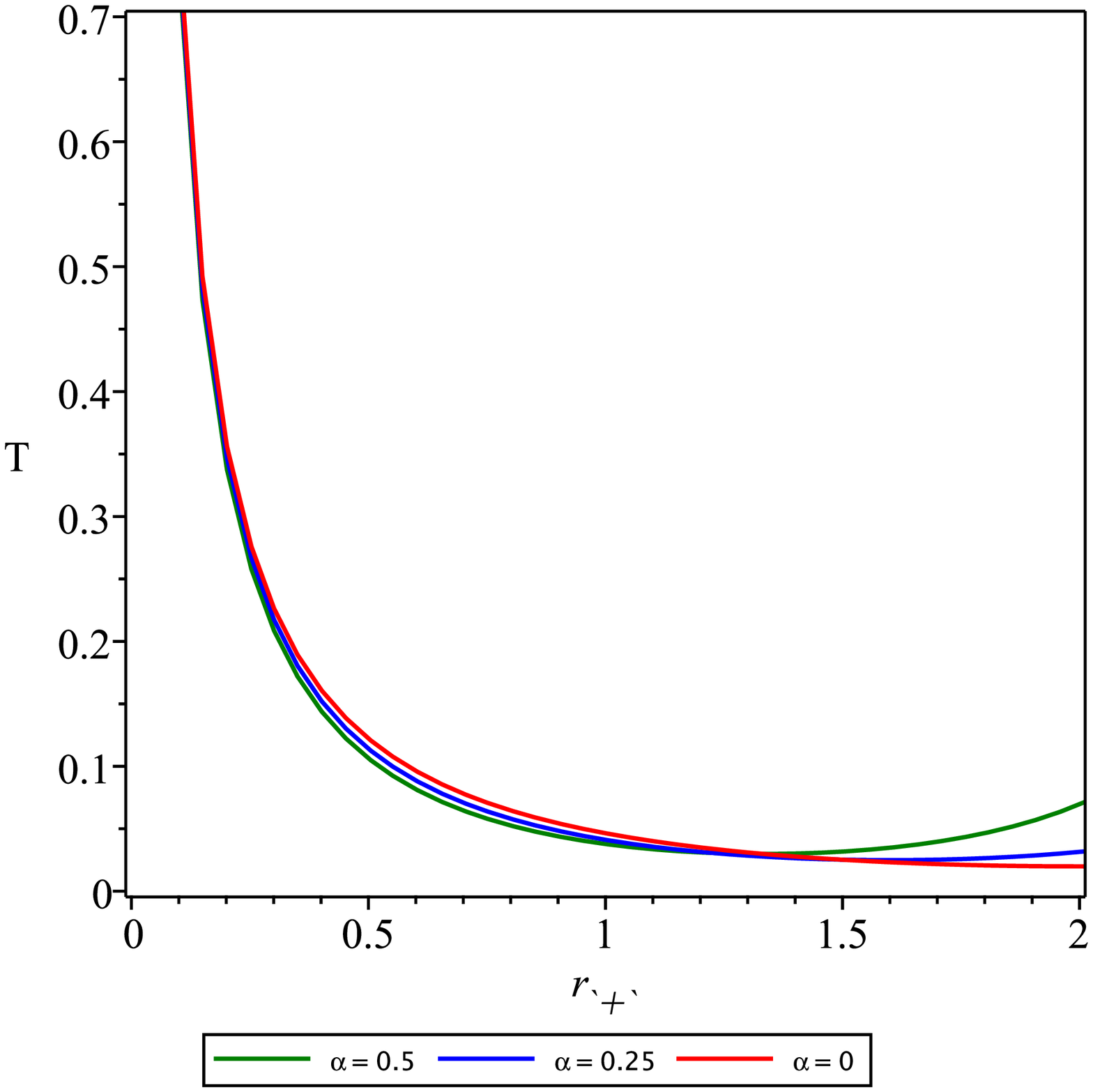}
	\includegraphics[width=0.5\textwidth]{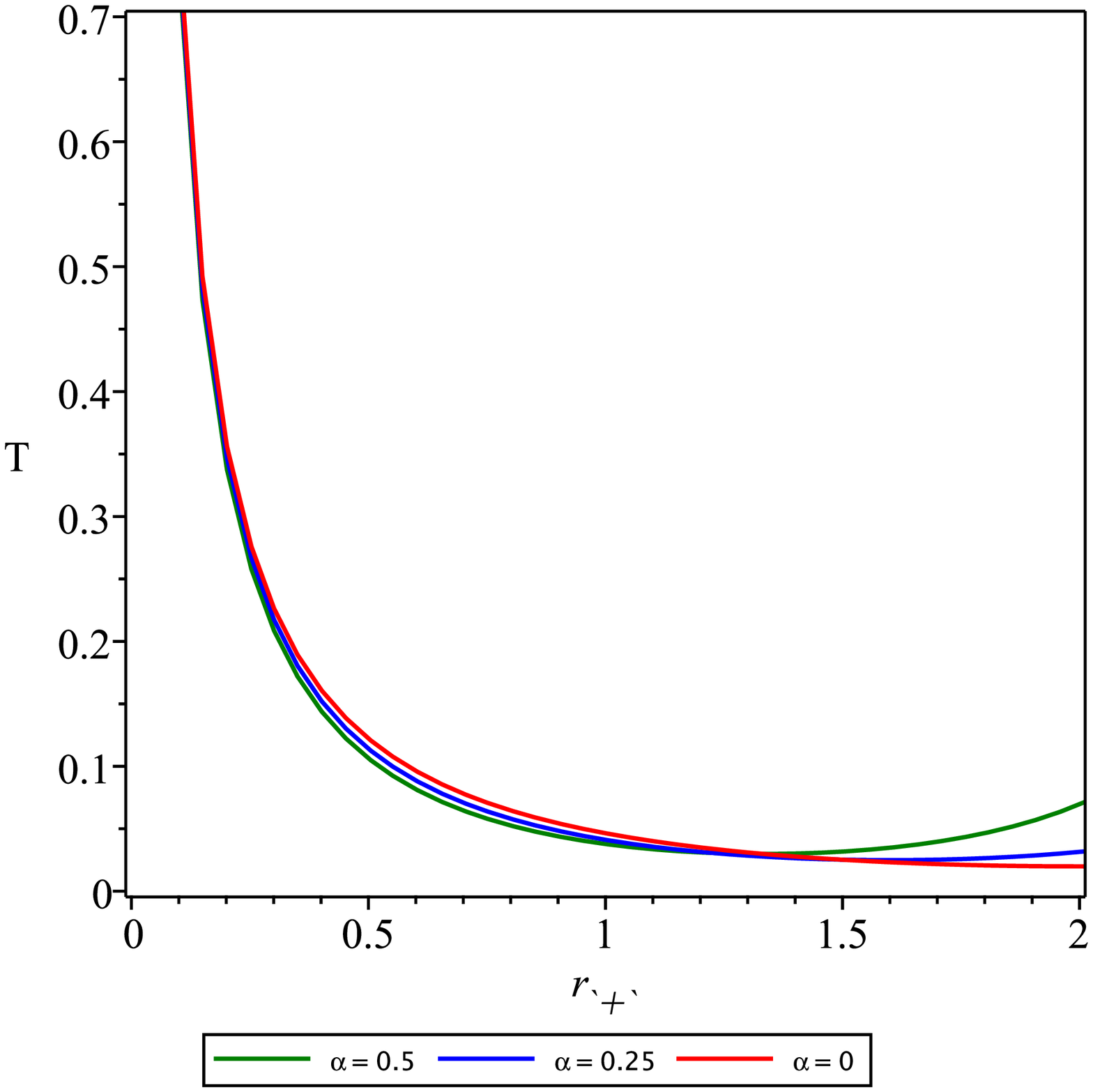}
	}
	\caption{The behaviour of the black hole temperature  against ${r_+}$, for different values of $\alpha$ and $\tilde{\gamma}$. In the two panels the results based on the Schwarzschild  type black hole are also presented. Here $G_N=1$ is assumed. For the left panel $\gamma=0.25$ and for the right panel $\alpha=0.5$.}
	\label{Pic:Temperature1-2}
\end{figure}

\begin{figure}[h]
	\centering
	\subfigure{
		\includegraphics[width=0.5\textwidth]{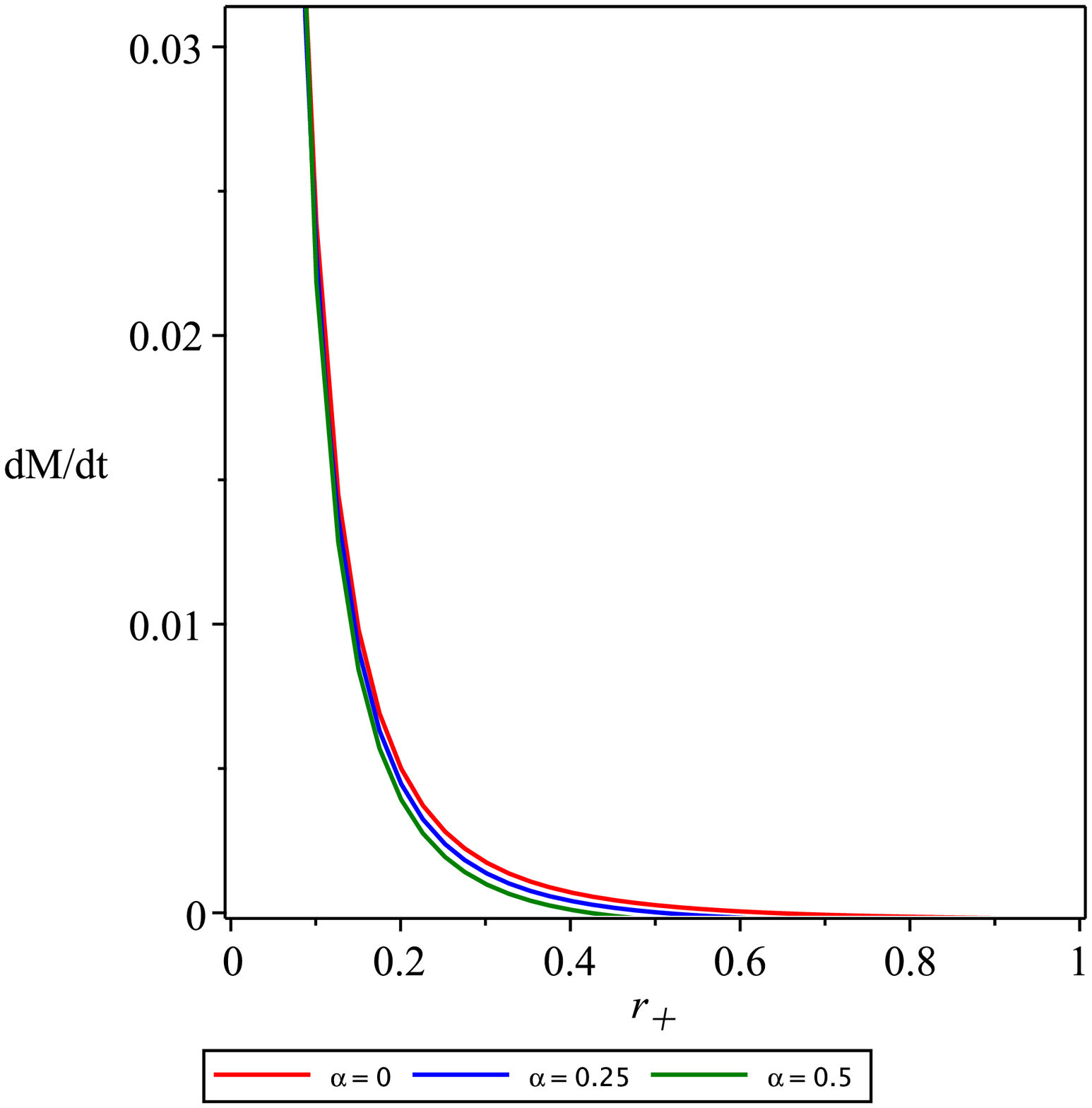}
	\includegraphics[width=0.5\textwidth]{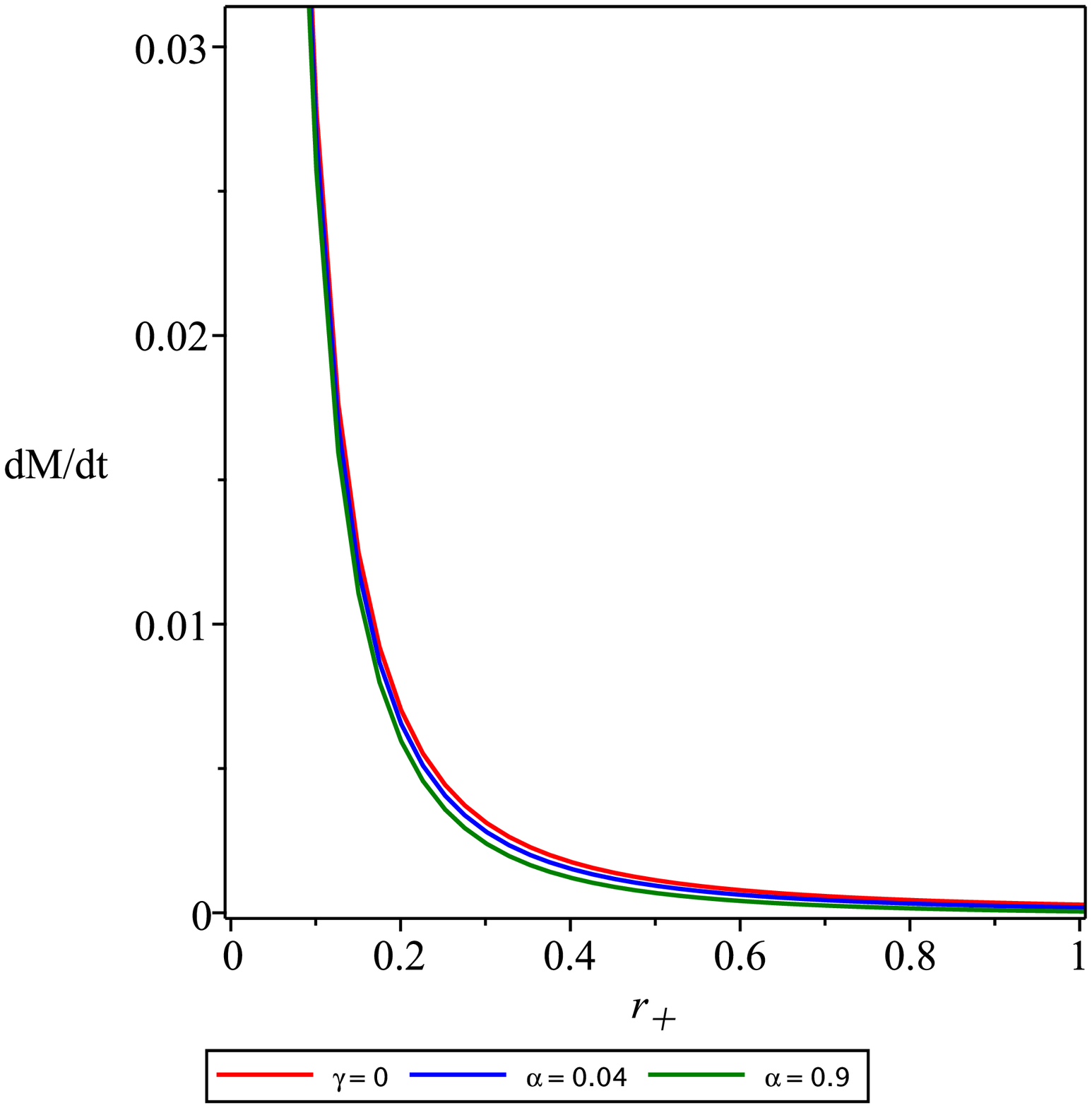}
	}
	\caption{The behaviour of the mass loss  against ${r_+}$, for different values of $\alpha$ and $\tilde{\gamma}$. In the two panels the results based on the Schwarzschild  type black hole are also presented. Here $G_N=1$ is assumed. For the left panel $\gamma=0.25$ and for the right panel $\alpha=0.5$.}
	\label{Pic:dMdt1-2}
\end{figure}

\begin{figure}[h]
	\centering
	\subfigure{
		\includegraphics[width=0.45\textwidth]{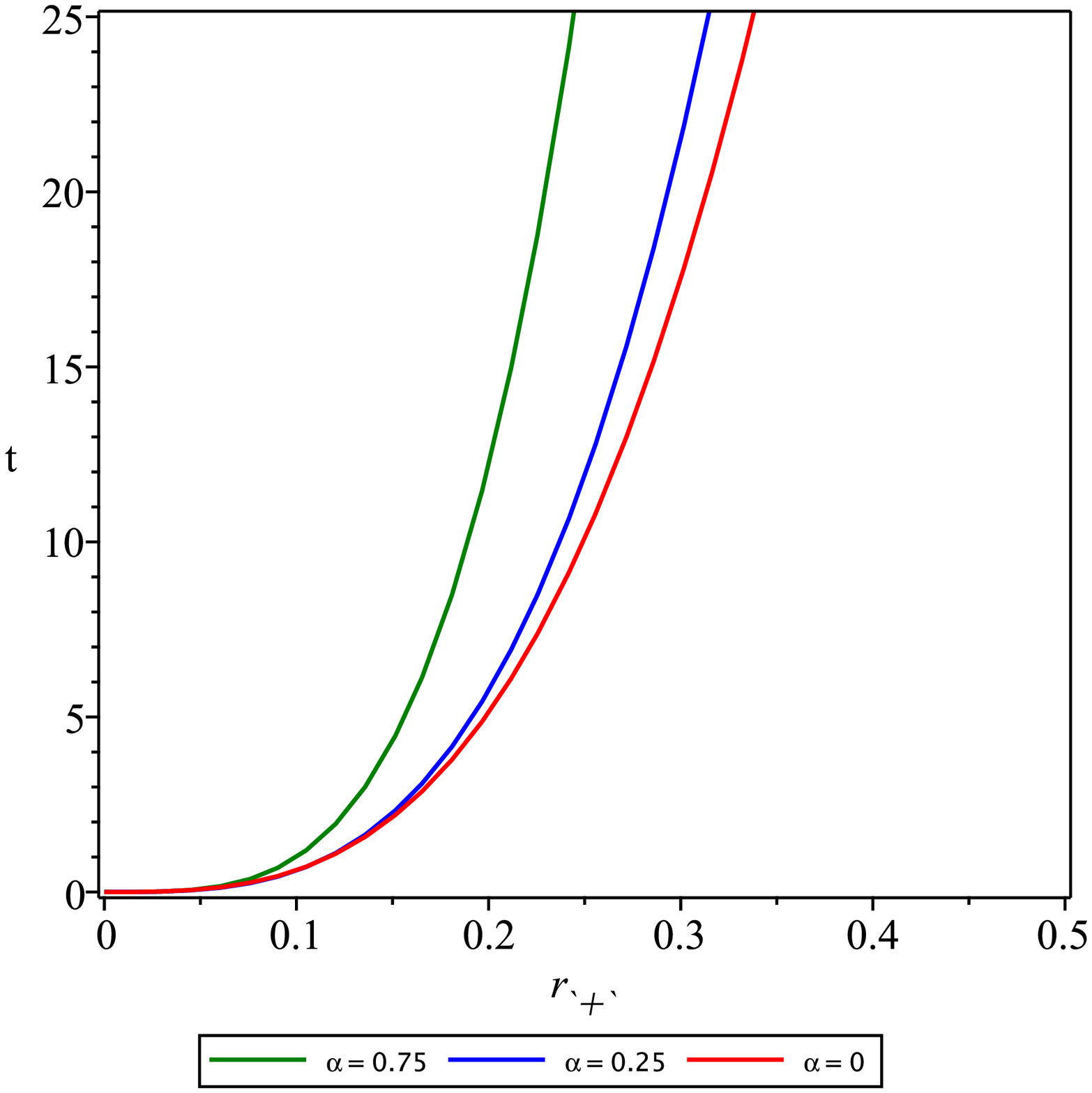}
	\includegraphics[width=0.45\textwidth]{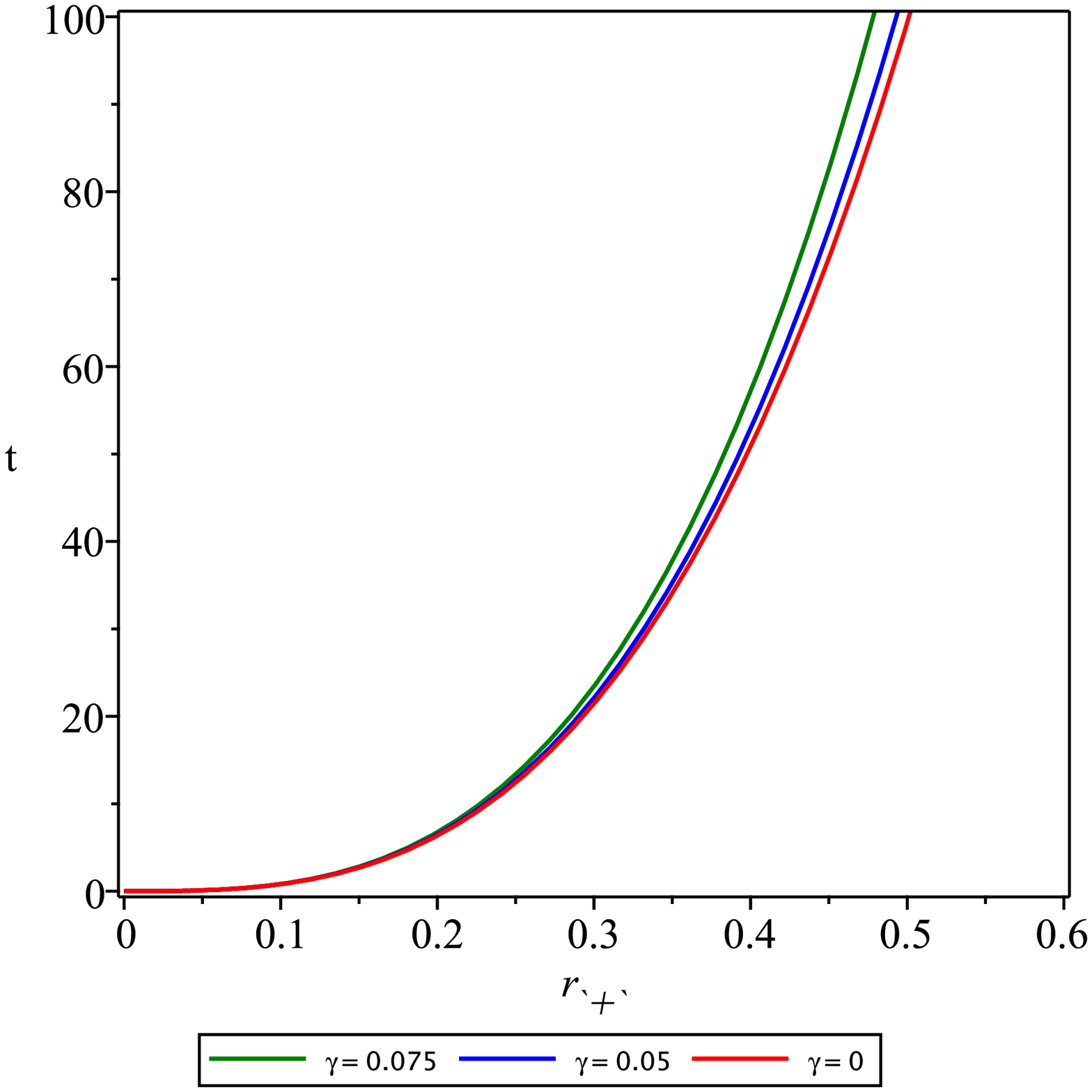}
	}
	\caption{The behaviour of the evaporation time  against ${r_+}$, for different values of $\alpha$ and $\tilde{\gamma}$. In the two panels the results corresponding to the Schwarzschild  type black hole are also presented. Here $G_N=1$ is assumed. For the left panel $\gamma=0.25$, while for the right panel $\alpha=0.5$.}
	\label{Pic:trr1-2}
\end{figure}

By substituting Eq.~\eqref{eqq4.14} into the equation \eqref{eqq4.12} for massless particles, i.e., for $\mu=0$, one  gets,
\begin{equation}
\sigma_{0}\sim 27\pi G^2 M^2+243\pi G^4M^3 \gamma+\mathcal{O}(\gamma^2)\,.
\end{equation}
Hence, the radius of the event horizon is obtained as,
\begin{equation}
r_{+}=\dfrac{1-\sqrt{1-8G^2\gamma M}}{2G\gamma}\,.
\end{equation}
For $\gamma$ small, satisfying the condition $\gamma\ll 1$, we obtain
\begin{equation}
r_{+}\sim 2GM+4G^3M^2\gamma+\mathcal{O}(\gamma^2)\,\sim \dfrac{2}{3}r_{c}+\dfrac{G\gamma}{9}r_{c}^{2}+\mathcal{O}(\gamma^2).
\end{equation}
Considering the above expression of the event horizon, the temperature of the black hole is given as follows,
\begin{equation}
T=\dfrac{f^{\prime}(r)}{4\pi}=\dfrac{1}{8\pi G M}-\dfrac{3G\gamma}{4\pi}+\mathcal{O}(\gamma^2)\,,
\end{equation}
where $^\prime$ denotes differentiation with respect to $r$. The behaviour of $T$ against $r_+$ is shown in Fig.\eqref{Pic:Temperature1-2}
 Finally, using this definition for the temperature, the mass loss rate reads,
\begin{equation}
\dfrac{dM}{dt}\sim \dfrac{27}{4096}\dfrac{\rho \varsigma}{\pi^3 G^2 M^2}-\dfrac{405}{4096}\dfrac{\rho  \varsigma \gamma}{\pi^3 M}+\mathcal{O}(\gamma^2)\,.
\end{equation}
where $ \varsigma=\pi^2/60.$ The behaviour of the mass loss rate  for different situations are illustrated in Fig.\eqref{Pic:dMdt1-2}.
By integrating the above expression one obtains the lifetime of the black hole as,
\begin{equation}
t=\dfrac{4096}{81}\dfrac{G^2\pi^3 M^3}{ \varsigma \rho}+\dfrac{5120}{9}\dfrac{G^4 \pi^3 M^4 \gamma}{\rho  \varsigma}+\mathcal{O}(\gamma^2)
\end{equation}
Here one should notice that, $G=G_N(1+\alpha)$. Beside this modification, the second term appears due to the presence of the quintessence scalar field surrounding the black hole. Here to realize the behaviour of the lifetime of the black hole one can consider the plots appear in the Fig.\eqref{Pic:trr1-2}

\section{Concluding remarks}\label{section6}

In the present work we have investigated in detail some astrophysical effects related to the possible presence of exotic forms of matter around black hole type cosmic objects. We have adopted, as a particular black hole model, the vacuum solution of the field equations in the MOG theory proposed, and developed, in \cite{Moffat:2005si}. This theory is a scalar-vector-tensor type theory, and its black-hole solutions has remarkable similarities with both the classic Reissner-Nordstr\"{o}m solution of general relativity, as well as with the Kiselev black hole solutions \cite{Kiselev:2002dx}, which initially interpreted as a black hole solution in the presence of a quintessence field. However, this interpretation is problematic \cite{Visser}, but the Kiselev solution, interpreted as a solution of the field equations in the presence of an exotic fluid, still has many interesting physical and astrophysical features that could lead to a better understanding of the properties of black holes embedded in an exotic cosmic environment. The solution of the field equations in the MOG gravity contains a term similar to the one appearing in the Reissner-Nordstr\"{o}m solution, describing the properties of a charged black hole, of the form $Q^2/r^2$, but with $Q$ unrelated to the electric charge, but proportional to the mass and to the effective gravitational constant of the theory.  Still, in order to point out to the similarities to the charged general relativistic black holes we call this class of vacuum solutions of the MOG theory as QMOG black holes.

The first important property of black holes is the position of their event horizon. The radius of the event horizon essentially depends on the parameter $\varepsilon$ of the equation of state of the exotic matter. In general, the position of the event horizon can be obtained for the QMOG black hole as a solution of a nonlinear algebraic equation. However, for some particular values of the parameter of the equation of state $\varepsilon$, some polynomial equations can be obtained.

The dynamical characteristics of the motion of massive test particles around a QMOG black hole can be obtained from the properties of the effective potential, which can be defined through the geodesic equations of motion. The effective potential also allows the determination of another important characteristic of the black hole geometry, the location of the innermost stable circular orbits. For the QMOG black hole, the positions of the event horizons and of the stable circular orbits depend on the solution parameters $\alpha$ and $\gamma$, as summarized in Tables \ref{tab:alpha} and \ref{tab:Q}, respectively. The effective potential is also strongly dependent on the solution parameters. for example, as one can see from Fig.~\ref{Pic:Veff}, the increase in the value of  $ \tilde{\gamma} $, leads to a decrease of the magnitude of the effective potential.

Three important effects that could help to discriminate between different black hole types are the light deflection, the Shapiro delay and the shadow of the black hole, respectively. We have investigated in detail these effects, and we have explicitly obtained, by using some algebraic approximations, the explicit expressions for the deflection angle, the angular radius and the diameter of the shadow, as well as the time delay due to the presence of the black hole.   With the help of each of these observations, one can obtain some constraints on the parameters of the black hole solution. As an interesting result, we found an upper bound on the quintessence parameter, $\gamma < 10^{-17}$. To obtain this upper bound we have supposed that the parameter $\alpha$ is much smaller than one in the Solar System tests. The light deflection angle for a black hole in this configuration was also obtained. Important information on the nature of the central compact object can be found through the study of its shadow. We have studeied in detail the shadow of the QMOG black hole, and obtained the angular radius, diameter, and shape of the  lack hole. These important information may provide the theoretical basis for constraining the parameters of the modified gravity model, and discriminate between the different types of black holes.

Many astrophysical objects interact gravitationally with the cosmic environment, and grow through matter accretion. The Universe is filled with interstellar matter, whose density was much higher in the early stages of cosmological evolution. The presence of interstellar matter determines the formation of accretion disks, located  around compact objects. Accretion disks are highly dynamical system, and in particular they are an important source of energy, emitted mostly in the electromagnetic spectrum. Neutrino emissivity of the disks is also possible. The emission of the electromagnetic radiation by the disk is mainly a consequence of the presence and influence of the external gravitational field  of the central compact object. The gravitational field of the massive star around which the disk is formed is determined by its nature - black hole, neutron or quark star,  naked singularity etc. Thus, the  direct observations of the electromagnetic emission spectra from astrophysical accretion disks give us the chance of obtaining essential information on the physical and astrophysical properties of the massive object around which the accretion disk was formed due to the interaction with the cosmic environment. Hence, modified gravity theories, can be tested and constrained by using the properties of their accretion disks. Moreover, one can obtain a large amount ofrelevant  astrophysical information from the observation of the motion of matter around the compact object.

In the present study we have performed a detailed analysis of the emissivity properties of the accretion disks that form around the QMOG black holes. By using the explicit form of the metric, the expression of the electromagnetic flux can be obtained in an exact analytical form. The flux is dependent on the parameters of the metric, and, for certain values of the parameters, it shows a significant difference as compared to the general relativistic Schwarzschild case. The differences do appear at the level of the maximum values of the flux, and in the position of the maximum. For the QMOG balck holes, there is a slight displacement of the maximum for higher $r$ values.  A similar pattern can be observed in the temperature distribution on the surface of the disk, with the temperature maximum displaced towards the interior of the disk. Similar differences do appear in the case of the efficiency parameter.

The quantum properties of the black holes are important for both theoretical, and astrophysical point of view. Even they are not directly detectable at the present moment, on a long run they can influence essentially the properties of the black holes. We have obtained the corrections to the Hawking temperature due to the extra terms present in the QMOG black hole solutions, and we have investigated its dependence on the model parameters. The mass loss rate and the evaporation time have also been obtained, and their dependencies on the model parameters have been investigated. Interestingly, an increase  in the lifetime of the black hole is obtained for the QMOG black holes. This is due to the presence of supplementary terms of the form $\dfrac{5120}{9}\dfrac{G^4 \pi^3 M^4 \gamma}{\rho \varsigma}$, which do appear besides the usual general relativistic terms.

In the present study we have considered some basic astrophysical properties of a particular modified gravity theory black hole. The obtained results may open some new perspectives in the observational testing of this type of objects, and in discriminating between different types of compact objects. Investigating these corrections for other type of black holes, for example, rotating black holes, can be the subject of future work.

\section*{Acknowledgements}
 H.S. is grateful to H. Firouzjahi for the constructive discussions they had on black holes and singularities. He also thanks J. Moffat for the discussions on a primary version of the paper. The work of TH is supported by a grant of the Romanian Ministry of Education and Research, CNCS-UEFISCDI, project number PN-III-P4-ID-PCE-2020-2255 (PNCDI III).

\end{document}